\documentclass[]{aa} 
\usepackage{graphicx}
\usepackage[symbol]{footmisc}
\usepackage{xcolor}
\usepackage{natbib}  
\usepackage{scalefnt}	
\usepackage[varg]{txfonts}
\usepackage[breaklinks]{hyperref}
\usepackage{breakurl}

\def\Cdou {$\rm ^{12}C$}
\def\Ctre {$\rm ^{13}C$}

\def\Cdt {$\rm ^{12}C /^{13}C$}
\def\lisisept {$\rm ^{6}Li /^{7}Li$}
\def\Ou  {\ion{O}{i}}

\def\Nau  {\ion{Na}{i}}

\def\Alu  {\ion{Al}{i}}

\def\Teff  {$T_\mathrm{eff}$}
\def\logg  {$\log g$}

\def\vt    {$\rm v_{t}$}
\def\kms   {$\rm km\,s^{-1}$}

\begin{document}
\title{Detailed analysis of a sample of field metal-poor N-rich dwarfs.
\thanks{Based on observations collected at the European Organisation for Astronomical Research in the Southern Hemisphere (Archives of programmes 090.B-0504(A) PI Chaname; 095.D-0504(A) PI Melendez; 076.B-0166(A) PI Pasquini; 067.D-0086(A) PI Gehren; 071.B-0529(A) PI Silva; 065.L-0507(A) PI Primas), and collected at the W. M. Keck Observatory, archive programme G401H, PI Melendez. One star was also observed at ESO with the spectrograph ESPRESSO, programme 107.22RU.001 PI Spite, and two stars were observed at the ``Observatoire de Haute Provence'' (Archives of programme 10A.PNPS.HALB, PI Halbwachs) and at the Narval spectrograph of the ``Observatoire du Pic du Midi'', programme L172N04, PI Spite.}
}

\author{
M. Spite \inst{1}\and 
F.~Spite \inst{1}  
\thanks{deceased on 2021.07.21 \url{https://www.observatoiredeparis.psl.eu/disparition-de-francois-spite.html?lang=en}} \and
E. Caffau \inst{1}\and
P. Bonifacio\inst{1}\and
P. Fran\c cois\inst{1}  
 }
\institute {
GEPI, Observatoire de Paris, PSL Research University, CNRS,
Place Jules Janssen, 92190 Meudon, France
}

\authorrunning{Spite M. et al.}
\titlerunning{Metal-poor N-rich dwarf stars}

\abstract
{}
{The aim of this work is to compare the detailed chemical composition of the field N-rich dwarf stars to the second generation stars of globular clusters (GC) in order to investigate the hypothesis that they originated in GCs.}
{We have measured the abundance of 23 elements (from Li to Eu)  in a sample of six metal-poor N-rich stars (three of them pointed out for the first time) and we have compared their chemical composition to, (i) the chemical composition observed in a sample of classical metal-poor stars, and (ii) the abundances observed in the second generation stars of GCs.
}  
{In metal-poor N-rich stars C and O are slightly deficient but the scatter of [(C+N+O)/Fe] is very small, a strong indication that the N enrichment is the result of a pollution by CNO processed material. The N-rich stars of our sample, like the second generation stars in the GCs, show an excess of Na and sometimes of Al, as expected if the material from which these stars were formed, has been polluted by the ejecta of massive AGB stars. For the first time we have been able to establish an anti-correlation Na-O in field stars like the one observed in NGC6752. The N-rich star HD\,74000 has a rather low [Eu/Ba] ratio for its metallicity. Such an anomaly is also observed in several second generation stars of M15.
}
{This analysis supports the hypothesis that the N-rich stars today observed in the field, were born as second generation stars in GCs. 
}

\maketitle
%
\section{Introduction}

It is well known that most globular clusters (GC) host multiple populations identified by their different chemical compositions. Compared to the first population, the second population is enriched in He, N and Na and depleted in C and O \citep{BastianLardo18,MucciarelliLL19,GrattonBC19}.\\

To date, there is no complete consensus on the origin of these multiple populations. 
Generally the second population stars is supposed to be the result of a pollution by the ejecta of first generation stars \citep[see ][]{BastianLardo18}: intermediate mass asymptotic giant branch (AGB) stars ($3<M<9M_{\odot}$),  fast rotating massive stars (FRMS) with $M > 15\,M_{\odot}$ \citep{MaederMeynet06},  and even very massive stars (VMS) with $ M\sim10^{4}M_{\odot}$ \citep{DenissenkovHartwick14}.
\\
Recently it has been proposed by \citet{GielesCharb19} that a very massive star with a mass
$ M>1000\,M_{\odot}$ forms via stellar collisions, during GC formation, and pollutes the intra-cluster medium.\\
Precise determinations of the chemical composition of the first and second populations have been obtained \citep[see for example:][]{PasquiniBM05,PasquiniBR07,PasquiniEB08,LindPC09}, and anti correlation between the abundance of N, Na and C, O has been found to be nearly universal among old and massive clusters \citep{GrattonBC19}.

Thanks to the Gaia parallaxes and proper motions
we have now direct evidence of stellar streams 
formed by stars lost by GCs \citep{IbataMM21}.
Some GCs show also extended tidal tails.
In some cases, like C-19, the narrow metallicity dispersion and sizeable
dispersion in Na abundances implies that a stream results
from the disruption of a GC, although the cluster, as such,
no longer exists \citep{MartinVA22,YuanMI22}.
Prior to the availability  of the Gaia data, the identification of
stars lost from GCs had to rely on their chemical properties.

Large systematic studies were undertaken to find field stars, presenting the same chemical anomalies as those observed in globular clusters   \citep[see e.g.][]{MartellSB11,RamirezMC12}. More recently N-rich metal-poor field giants have been discovered in the Galactic bulge thanks to the large spectroscopic survey APOGEE \citep{SchiavonJF17}  and also in the field of our Galaxy with LAMOST \citep{TangFTL20}. Their analysis suggests that the origin of these stars is indeed related to the disintegration of GCs. However such an origin is discussed and \citet{Bekki19} suggests that N-rich stars in the Bulge and in the halo can be formed from the destruction of high-density building blocks where nitrogen-rich stars are formed from an interstellar medium heavily polluted by AGB ejecta.

In the Galaxy, at least, three classes of metal-poor N-rich stars may be found.\\ 
i--  The metal-poor giant stars after the bump undergo a deep mixing which brings the CNO processed material to the surface \citep{GrattonSC00,SpiteCP05,SpiteCH06} and the stars appear to be C-poor and N-rich.
\\ 
ii-- At low metallicity, many stars, dwarfs or giants, are N-rich and C-rich. Following \citet{LucatelloBC06}, more than 20\% of the metal-poor stars with $ \rm[Fe/H] \leq -2.0$ belong to this class of Carbon enhanced metal-poor (CEMP) stars.  The peculiar chemical composition of these stars, rich in C {\it and} in N, and often also in neutron-captured elements, is  generally explained by pollution by different kinds of AGB stars during the life of the observed stars or by a formation from a material ejected by rotating massive stars \citep[e.g.][]{MasseronJP10,HansenAN16b,HansenAN16a,ChoplinMM16,ChoplinHM17}. Another possibility, invoked in particular, to explain the CEMP stars not enriched in neutron-capture elements (CEMP-no), is an enrichment of the cloud that formed the star by faint supernovae providing only carbon and the lighter elements \citep{IshigakiTK14,BonifacioCS15}.
\\ 
iii-- A third class of stars,  the one we are interested in, in the present paper, contains N-rich stars, with a normal C abundance. This type of stars seems to exist among field dwarf and giant stars. It is of particular interest to detect dwarf stars belonging to this class, since their atmosphere has not undergone any mixing with CNO processed material of the deep layers. It is then more secure to compare their chemical composition to the chemical composition of dwarfs in GCs. 

\begin{table*}
\caption[]{Spectral ranges ( in nm) of the spectra used in this study. The UVES spectrograph is installed at the ESO-VLT, HARPS at the ESO 3.6m telescope, FEROS at the 2.2m ESO telescope, SOPHIE at the 1.93m telescope of the ``Observatoire de Haute Provence'' (OHP), HIRES at the 10m Keck telescope and ESPRESSO at the incoherent combined Coud\'e facility of the ESO-VLT. }
\label{Tab:spectra}
\scalefont{0.85}
\centering
\begin{tabular}{l@{~}c@{~~}c@{~~}c@{~~}c@{~~}c@{~~}c@{~~}c@{~~}c@{~~}c@{~~}c@{~~}c@{~~}}
\hline
             &           &UVES-B390 or&           &           &         &         &       & HIRES  & HIRES  & HIRES   \\
  Star       &UVES-B346  &UVES-B437   & UVES-R580 & UVES-R860 & HARPS   & FEROS   &SOPHIE & CCD1   & CCD2   & CCD3   & ESPRESSO \\
\hline
HD\,25329&      -        &     -      &    -      &    -      &   -     &   -     &  -    &302-390 &393-489 &494-580  \\
~~~~~~'' &      -        &     -      &    -      &    -      &   -     &   -     &  -    &   -    &594-750 &753-900  \\
HD\,74000&    305-387    &     -      &    -      & 671-1000  & 378-690 &    -    &  -    &   -   \\
HD\,97916&    308-387    &     -      &478.5-680  &    -      &    -    &   -     &387-694&   -   \\
HD\,160617&   305-387    &     -      &    -      &    -      & 378-690 & 390-910 &  -    &   -   \\
HD\,166913&   307-387    &     -      &478.5-680  &    -      &    -    &    -    &  -    &   -    &        &        & 370-787\\
G24-3&        306-387    &  377-498   &478.5-680  & 672-1000  &    -    &    -    &  -    &307-399\\
G53-41&          -       &  329-452   &478.7-680  &    -      &    -    &    -    &  -    &   -   \\
G90-3&           -       &     -      &     -     &    -      &    -    &    -    &  -    &302-394& 391-485 & 491-580  \\
\hline
\end{tabular}
\end{table*}

In metal-poor giant stars, the nitrogen abundance can be measured in the visible or in the red from the CN bands \citep{SchiavonJF17,TangFTL20} but in metal-poor dwarfs the lines of these bands are too weak and only the vibrational band of NH in the near UV (336\,nm) can be used \citep[see e. g. ][]{IsraelianER04}. As a consequence, very few N-rich metal-poor dwarfs have been detected so far. As far as we know, only five metal-poor dwarfs with $ \rm [Fe/H] \leq -1.0$, have been reported to have a ratio [N/Fe]>0.6: HD\,25329, HD\,74000, HD\,97916, HD\,160617, and HD\,166913 by in particular \citet{HarmerPagel73}, \citet{Schuster81}, \citet{Laird85}, \citet{CarbonKB86},  \citep[see also][]{SpiteSpite86}.

 In the present paper we analyse in detail these five stars suspected (or already known) to be N-rich, and we report the existence of three more metal poor dwarfs strongly enriched in nitrogen ($ \rm[N/Fe] \geq 1.0$). In most of these stars we could measure the abundance of the elements extended from Li to Eu, and compare the abundance pattern of the elements with the abundance pattern of normal metal-poor stars and of the second generation stars in GCs. 

\begin{figure}
\label{Fig:iso1}
\begin{center}
{\includegraphics [scale=0.6]{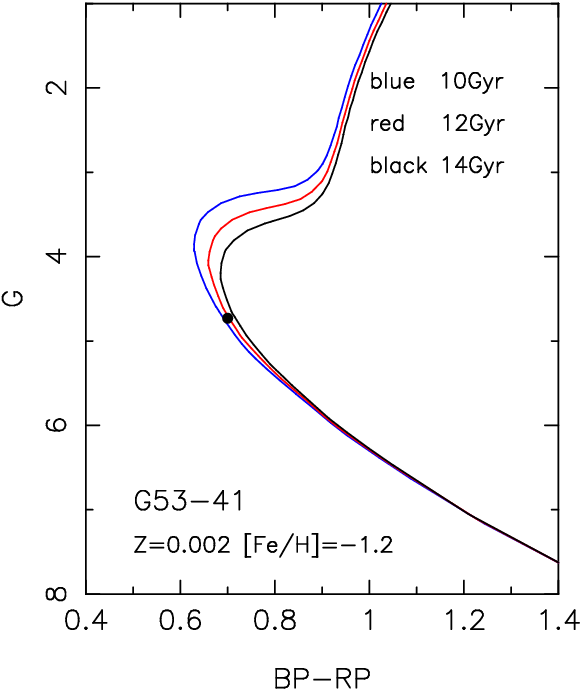}}
\end{center}
\caption[]{Position of G53-41 (black dot) in a G vs BP-RP diagram, compared to PARSEC isochrones computed for 10, 12 and 14 Gyr.}
\end{figure}

\section{Star sample and reduction}
Near UV spectra of metal-poor dwarf stars around 336\,nm, have been retrieved from the ESO-VLT-UVES and Keck-HIRES archives.  These spectra have been often obtained in the aim of studying the behaviour of Be in these stars and the parameters of their model are, at least in a first approximation, already known. Among these stars we visually selected three new dwarfs or turnoff stars presenting a very strong NH band:  G24-3, G53-41, G90-3. Following  \citet{RamirezMC12}, G53-41 is oxygen-poor and Na-rich, key signatures of the abundance anomalies observed in GCs, and thus it was already suspected to be born in a GC.\\
Finally a sample of eight metal-poor dwarfs has been selected, the five stars already known or suspected to be N-rich and three new stars selected for their strong NH band in high resolution spectra.\\
The characteristics of all the spectra used in this analysis are given in  Table \ref{Tab:spectra}. All these spectra have a resolving power $R \geq 40000$ and the S/N in the region of the NH band larger than 100.

\begin{figure*}
\begin{center}
\resizebox{18.5cm}{!}
{\includegraphics [clip=true]{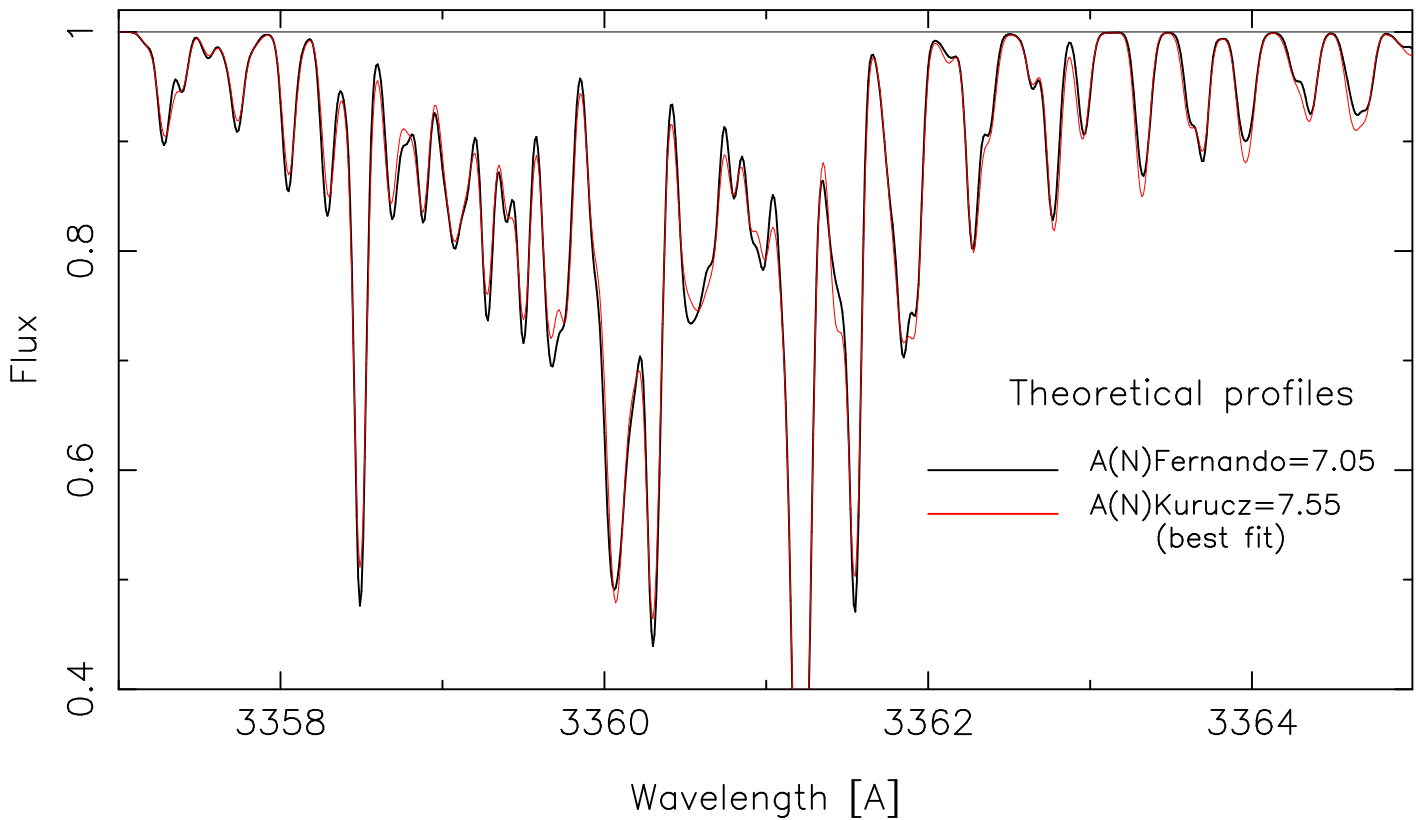}}
\end{center}
\caption[]{Fit of the theoretical NH band computed from the \citet{FernandoBH18} data (black line) with A(N)=7.05 and the parameters of HD\,74000, with synthetic spectra computed from Kurucz data. The best fit in the total interval 3357-3365\,\AA~ is obtained with A(N)=7.55 (red line). In this region, the mean difference of the N abundance computed with these two different set of data is thus 0.50.
}
\label{Fig:compNHband}
\end{figure*}

\begin{table}
\caption[]{Gaia DR2 data, photometry and model parameters of the stars}
\label{Tab:param}
\scalefont{0.75}
\centering
\begin{tabular}{l@{~}c@{~~~~~}c@{~~~}c@{~~~}c@{~~~}c@{~~}c@{~~}c@{~~}c@{~~}}
\hline
\noalign{\smallskip}
                 &HD      &HD     &HD     & HD    & HD    & G     &G      &G     \\
                 &25329   &74000  &97916  &160617 &166913 & 24-3  &53-41  &90-3  \\
\hline
\noalign{\smallskip}
dist (pc)        & 18.5   & 111   & 97    & 104   & 60    & 129   & 165   & 265  \\
\\
E(B-V)Stil       & 0.001  & 0.002 & 0.006 & 0.006 & 0.006 & 0.038 & 0.013 & 0.015\\
$ \rm G_{0}$     & 6.89   & 4.28  & 4.12  & 3.46  & 4.17  & 4.63  & 4.73  & 3.15 \\
$\rm(BP-RP)_{0}$ & 1.106  & 0.639 & 0.599 & 0.691 & 0.659 & 0.661 & 0.698 & 0.730\\
Age (Gyr)        & -      & 14    & 5     & 14    & 14    & 14    &  12   &  14  \\
\\
\Teff            & 4870   & 6260  & 6400  & 6000  & 6230  & 6030  & 6050  & 5900 \\
\logg            & 4.75   & 4.3   & 4.35  &  3.9  &  4.2  &  4.4  &  4.4  & 3.7  \\
\vt~(\kms)       & 0.8    & 1.2   & 1.2   &  1.2  &  1.3  &  0.9  &  0.9  & 1.2  \\
 $\rm[Fe/H]$     &-1.72   &-2.00  &-0.75  &-1.80  &-1.50  &-1.58  &-1.19  &-2.18 \\ 
\hline
\end{tabular}
\end{table}

\subsection{Model Parameters and ages}  \label{sec:Par}
We took advantage of the information provided by the Gaia DR2 \citep{GaiaDR2-Brown18,ArenouLB18} to derive new stellar parameters: effective temperature and surface gravity. We used the distance-dependent reddening provided by Stilism \footnote{\url{https://stilism.obspm.fr}} \citep{LallementBV18}, to deduce the absolute magnitude $ \rm G_{0}$ and the colours $ \rm(BP-RP)_{0}$ of the stars. We then compared these values to PARSEC isochrones\footnote{\url{http://stev.oapd.inaf.it/cgi-bin/cmd}} \citep{BressanMG12,MarigoGB17} and we derived \Teff, \logg~ and the age of the star. 
The data are given in Table \ref{Tab:param}, and an example of the isochrones is shown in Fig.\,\ref{Fig:iso1}. 
The distance of the stars has been computed by a simple inversion of the parallax after correcting by the zero-point offset of 0.03 \citep{LindegrenHB18}. Since the stars are close, the error on the distance of the star is always less than 3\,pc, the error on  $ \rm G_{0}$ less than 0.06, and the error on $\rm(BP-RP)_{0}$ less than 0.03. We finally estimate that the error on \Teff~ is less than 100K and the error on \logg~ less than 0.2\,dex. 
The age of all the stars is between 12 and 14\,Gyr (Table \ref{Tab:param}), with two exceptions:\\
i--We could not measure the age of HD\,25329 because it is a rather cool star and still on the main sequence, but from a Bayesian estimation \citet{Pace13} reports for this star, known to have a strong chromospheric activity, an age of about 5\,Gyr,  surprising for a star known to have a metallicity lower than [Fe/H]=--1.7.\\
ii--The age of HD\,97916 ([Fe/H]=--0.75) seems also close to 5\,Gyr (Table \ref{Tab:param}).\\
These two stars will be discussed hereafter.

\begin{table*}
\caption[]{Abundance of the elements. As usual, A(X)= 12 + log (N(X)/N(H)),  [X/H]=$A(X)_{*}-A(X)_{\odot}$ and [X/Fe]=[X/H]--[Fe/H]. The six N-rich stars are at the top of the table, followed by the two stars (HD97916 and HD166913) that were found to have a normal N abundance. The adopted solar abundances are in bold type.}
\label{Tab:abundNH}
\scalefont{0.75}
\centering
\begin{tabular}{l@{~~}c@{~~~~~~~}c@{~}c@{~~~~~}c@{~}c@{~~~~~}c@{~}c@{~~~~~}c@{~}c@{~~~~~}c@{~}c@{~~~~~}c@{~}c@{~~~~~}c@{~}c@{~~~~~}c@{~}c@{~~~~~}c@{~}c@{~~~~~}c@{~}c@{~~~~~}c@{~}c@{~~~~~~}c@{~}}
\hline
~~       &          &         &         &         &         &         &         &         &        &         &         & adopted&adopted &         &         &         &       \\
   Ident.& [Fe/H]   & A(Li)   &  A(Be)  &A(C)$\rm_{CH}$& [C/Fe]&A(N)$\rm_{NH}$&[N/Fe]&A(O)$_{777}$&[O/Fe]&A(O)$\rm_{OH}$&[O/Fe]&A(O)&[O/Fe]  &A(Na)&[Na/Fe]  &  A(Mg)  &[Mg/Fe]\\
{\bf Sun}&          &         &         &{\bf8.50}&         &{\bf7.86}&         &{\bf8.76}&         &{\bf8.76}&   &{\bf8.76}&      &{\bf6.30}&         &{\bf7.54} &            \\
 HD 25329&    -1.72 & -       &   -     &    7.08 &    0.30 &    7.34 &    1.20 &    7.47 &    0.43 &    7.40 &0.36&   7.43 & 0.39 &    4.69 &    0.11 &    6.03 &    0.21 \\
 HD 74000&    -2.00 & 2.28    &  -0.40  &    6.68 &    0.18 &    7.05 &    1.19 &    7.23 &    0.47 &    7.30 &0.54&   7.28 & 0.52 &    4.50 &    0.20 &    5.87 &    0.33 \\
HD 160617&    -1.80 & 2.32    &  -0.30  &    6.66 &   -0.04 &    7.08 &    1.02 &    7.36 &    0.40 &    7.40 &0.44&   7.38 & 0.42 &    4.51 &    0.01 &    5.99 &    0.25 \\
    G24-3&    -1.58 & 2.12    &  -0.02  &    6.63 &   -0.29 &    7.40 &    1.12 &    7.52 &    0.34 &    7.38 &0.20&   7.44 & 0.26 &    4.95 &    0.23 &    6.16 &    0.20 \\
   G53-41&    -1.19 & 2.17    &    -    &    7.02 &   -0.29 &    7.74 &    1.07 &       - &    0.18*&    7.60 &0.03&    -   & 0.18 &    5.16 &    0.05 &    6.55 &    0.20 \\
    G90-3&    -2.18 & 2.30    &  -0.61  &    6.50 &    0.18 &    7.02 &    1.34 &       - &       - &    7.11 &0.53&   7.11 & 0.53 &    4.39 &    0.27 &    5.74 &    0.38 \\
~~       & \\         
HD 97916 &    -0.75 &  -      &  <-0.7   &   7.53 &   -0.22 &    6.91 &   -0.20 &    8.80 &    0.79 &    8.20 &0.19&   8.52 & 0.51 &    5.35 &   -0.20 &    6.88 &    0.09 \\
HD 166913&    -1.50 & 2.46    &   0.45  &    7.22 &    0.22 &    6.17 &   -0.19 &    7.81 &    0.45 &    7.92 &0.66&   7.82 & 0.56 &    4.54 &   -0.26 &    6.40 &    0.36 \\
\hline 
~~       &  \\

   Ident.& [Fe/H]   &   A(Al) & [Al/Fe] &  A(Si)  &[Si/Fe]  &  A(Ca)  &[Ca/Fe]  &  A(Sc)  &[Sc/Fe]  &  A(Ti)  &[Ti/Fe]  & A(Cr1) &[Cr/Fe] & A(Mn)  &[Mn/Fe]  & A(Co)  &[Co/Fe]\\
{\bf Sun}&          &{\bf6.47}&         &{\bf7.52}&         &{\bf6.33}&         &{\bf3.10}&         &{\bf4.90}&        &{\bf5.64}&       &{\bf5.37}&        &{\bf4.92}&       \\
 HD 25329&    -1.72 &    5.00 &    0.25 &    6.36 &   0.56  &    4.91 &    0.30 &    1.55 &    0.17 &    3.59 &    0.41 &   3.92 &   0.00 &   3.37 &   -0.28 &   -    &   -   \\
 HD 74000&    -2.00 &    4.35 &   -0.12 &    5.83 &   0.31  &    4.66 &    0.33 &    1.30 &    0.20 &    3.35 &    0.45 &   3.56 &  -0.08 &   3.10 &   -0.27 &   3.11 &  0.19 \\
HD 160617&    -1.80 &    4.59 &   -0.08 &    6.02 &   0.30  &    4.83 &    0.30 &    1.50 &    0.20 &    3.50 &    0.40 &   3.79 &  -0.05 &   3.47 &   -0.10 &   3.30 &  0.18 \\
    G24-3&    -1.58 &    4.70 &   -0.19 &    6.21 &   0.27  &    5.05 &    0.30 &    1.71 &    0.19 &    3.70 &    0.38 &   3.97 &  -0.09 &   3.64 &   -0.15 &   3.56 &  0.22 \\
   G53-41&    -1.19 &    4.94 &   -0.34 &    6.57 &   0.24  &    5.35 &    0.21 &    2.10 &    0.19 &    4.04 &    0.33 &   4.38 &  -0.07 &   4.25 &    0.07 &   4.08 &  0.35 \\
    G90-3&    -2.18 &    4.62 &    0.33 &    5.75 &   0.41  &    4.45 &    0.30 &    1.10 &    0.18 &    3.15 &    0.43 &   3.35 &  -0.11 &   2.92 &   -0.27 &   3.01 &  0.27 \\
~~       &  \\

 HD 97916&    -0.75 &    5.35 &   -0.37 &    6.74 &  -0.03  &    5.80 &    0.22 &    2.65 &    0.30 &    4.54 &    0.39 &   4.83 &  -0.06 &   4.30 &   -0.32 &   4.54 &  0.37 \\
HD 166913&    -1.50 &    4.71 &   -0.26 &    6.45 &   0.43  &    5.18 &    0.35 &    1.78 &    0.18 &    3.86 &    0.46 &   4.08 &  -0.06 &   3.54 &   -0.33 &   3.58 &  0.16 \\
\hline
~~       &  \\
   Ident.& [Fe/H]   &  A(Ni)  &[Ni/Fe]  & A(Zn)  &[Zn/Fe] &  A(Sr)  &[Sr/Fe]  &  A(Y)  &[Y/Fe]  &  A(Zr)  &[Zr/Fe]  &  A(Ba) &[Ba/Fe] &  A(Eu) &[Eu/Fe]  \\
{\bf Sun}&          &{\bf6.23}&        &{\bf4.62}&        &{\bf2.92}&        &{\bf2.21}&       &{\bf2.58} &        &{\bf2.17}&       &{\bf0.52}&         \\
HD 25329 &    -1.72 &    4.57 &    0.06 &   3.03 &   0.13 &    1.52 &  0.32   &  0.69  &  0.20  &   1.55  &   0.69  &  0.78  &  0.33  & -0.35  &  0.85   \\
HD 74000 &    -2.00 &    4.23 &    0.00 &   2.73 &   0.11 &    1.36 &  0.44   &  0.28  &  0.07  &   1.05  &   0.47  &  0.32  &  0.15  & -1.38  &  0.10   \\
HD 160617&    -1.80 &    4.45 &    0.02 &   2.84 &   0.02 &    1.34 &  0.22   &  0.30  & -0.11  &   1.04  &   0.26  &  0.63  &  0.26  & -0.76  &  0.52   \\
G24-3    &    -1.58 &    4.66 &    0.01 &   3.05 &   0.01 &    1.51 &  0.17   &  0.63  &  0.00  &   1.29  &   0.29  &  0.79  &  0.20  & -0.92  &  0.14   \\
G53-41   &    -1.19 &    4.97 &   -0.07 &   3.38 &  -0.05 &    1.95 &  0.22   &  1.20  &  0.18  &   1.90  &   0.51  &  1.65  &  0.67  & -0.21  &  0.46   \\
G90-3    &    -2.18 &    3.91 &   -0.14 &   2.58 &   0.14 &    0.97 &  0.23   & -0.14  & -0.17  &   0.64  &   0.24  & -0.14  & -0.13  & -1.32  &  0.34   \\
~~\\
HD 97916 &    -0.75 &    5.55 &    0.07 &   3.81 &  -0.06 &    2.20 &  0.03   &  1.32  & -0.14  &   1.95  &   0.12  &  1.57  &  0.15  & -0.18  &  0.05   \\
HD 166913&    -1.50 &    4.82 &    0.09 &   3.11 &  -0.01 &    1.80 &  0.38   &  0.79  &  0.08  &   1.43  &   0.35  &  0.85  &  0.18  & -0.64  &  0.34   \\
~~\\
\multicolumn{6}{l}{*~ oxygen abundance following \citet{RamirezMC12} }\\
\hline
\end{tabular}
\end{table*}

\begin{table*}
\caption[]{Parameters of the model adopted for the comparison stars and abundances of the elements relative to iron.}
\label{Tab:abundNor}
\scalefont{0.75}
\centering
\begin{tabular}{lrrrrrrrrrrrrrrrrrrr}
\hline
 Star  & \Teff~\logg~\vt~& [Fe/H]& [C/Fe]& [N/Fe]& [O/Fe]$\rm_{777}$&[O/Fe]$\rm_{OH}$&[O/Fe]$\rm_{adopt}$& [Na/Fe]& [Mg/Fe]&[Al/Fe]&[Si/Fe]&[Ca/Fe]&[Sc/Fe]\\
HD19445&  6070~ 4.4~ 1.3& -2.15&     0.48&  -0.11&              0.64&            0.93&  0.79&                -0.08&    0.51&  -0.14&   0.55&   0.41&   0.32\\
HD76932&  6000~ 4.1~ 1.3& -0.94&     0.05&  -0.01&              0.68&            0.52&  0.57&                -0.04&    0.44&  -0.10&   0.33&   0.27&   0.34\\
HD84937&  6300~ 4.0~ 1.3& -2.25&     0.39&   0.09&              0.65&            0.78&  0.72&                -0.10&    0.30&  -0.33&   0.38&   0.47&   0.23\\
HD94028&  6050~ 4.3~ 1.2& -1.40&     0.10&  -0.29&              0.42&            0.47&  0.45&                -0.10&    0.32&  -0.23&   0.41&   0.23&   0.23\\
HD140283& 5750~ 3.7~ 1.4& -2.57&     0.46&   0.07&              0.90&            0.76&  0.78&                -0.29&    0.26&  -0.39&   0.38&   0.27&   0.10\\
~~\\ 
 Star  & \Teff~\logg~\vt&[Ti/Fe]&  [Fe/H]& [Cr/Fe]& [Mn/Fe]&[Co/Fe]&[Ni/Fe]&[Zn/Fe]&[Sr/Fe]& [Y/Fe]&[Zr/Fe]& [Ba/Fe]&[Eu/Fe]   \\ 
HD19445&  6070~ 4.4~ 1.3&   0.51&   -2.15&   -0.03&   -0.10&   0.33&   0.06&   0.18&   0.39&   0.06&   0.38&   -0.10&   0.37   \\ 
HD76932&  6000~ 4.1~ 1.3&   0.42&   -1.00&    0.01&   -0.05&   0.29&   0.17&   0.05&   0.30&  -0.02&   0.28&    0.10&   0.36   \\ 
HD84937&  6300~ 4.0~ 1.3&   0.45&   -2.25&    0.05&   -0.27&   0.14&  -0.07&   0.16&   0.39&  -0.01&   0.31&   -0.22&   0.38   \\ 
HD94028&  6050~ 4.3~ 1.2&   0.37&   -1.40&   -0.10&   -0.25&   0.19&   0.11&  -0.03&   0.36&   0.09&   0.37&    0.17&   0.15   \\ 
HD140283& 5750~ 3.7~ 1.4&   0.33&   -2.57&    0.08&   -0.29&   0.29&   0.12&   0.25&  -0.18&  -0.42&  -0.07&   -0.81&  -0.28   \\
\hline
\end{tabular}
\end{table*}

\subsection{Chemical composition of the stars}   \label{Sec:ab1}
We carried out a classical LTE analysis of the stars using MARCS model atmospheres \citep{GustafssonBE75,GustafssonEE03,Plez08} and the {\tt turbospectrum} spectral synthesis code \citep{AlvarezP98,Plez-code12}. The microturbulence velocity was derived from the Fe\,I lines, requiring that the abundance derived for individual lines be independent of the equivalent width of the line. The abundances of the elements and [Fe/H],
 are given in Table \ref{Tab:abundNH}. Note that in a first time we selected N-rich stars with [Fe/H]<--1.0, but with the new determination of the atmosphere parameters, HD\,97916 was found to be less metal-poor ([Fe/H]=--0.75]). However we decided to keep this star in our sample.

\subsubsection{Atomic data, and NLTE corrections}
  
The atomic data were mainly taken from the ``Vienna Atomic Line Database'' VALD3  \footnote{\url{http://vald.oreme.org/~vald/php/vald.php?}}. 
For Ba and Eu the isotopic and hyperfine structure of the lines has been taken into account. \\
In these dwarf stars the NLTE effects are generally rather weak and since the aim of this work is to compare the relative abundances in N-rich dwarfs to those in normal stars with very similar models, these effects were taken into account only for \Ou, \Nau~ and \Alu.\\ 
--In this type of stars, the oxygen abundance can be generally determined from the red oxygen permitted triplet at 777\,nm. Following \citet{ZhaoMY16} when $\rm -2.5 \leq [Fe/H] \leq -1.0$, the NLTE correction is small and close to about --0.1\,dex. This correction was applied in Table \ref{Tab:abundNH}.\\
--Our determination of the Na abundance is based on the secondary \Nau~lines at 5682, 5688, 6154 and 6160\,\AA. The NLTE correction has been estimated from \citet{LindAB11}. In all our stars it is very close to -0.1\,dex and this correction was applied in  Table \ref{Tab:abundNH}.\\
--To determine the Al abundance we had to use the the resonance lines of \Alu. The NLTE correction has been computed by \citet{NordlanderLind17} for different atmospheric models, for the turnoff stars the correction is large, (about +0.4\,dex), and it was applied in Table \ref{Tab:abundNH}.

\subsubsection{Molecular data}
The carbon abundance was determined from the CH band at 314.4\,nm and between 430 and 432.5\,nm. The abundance deduced from these two regions are in excellent agreement.
The parameters of the CH molecular bands \citep{MasseronPVE14} were directly taken from the Bertrand Plez site\footnote{\url{https://nextcloud.lupm.in2p3.fr/s/r8pXijD39YLzw5T}}.

The nitrogen abundance was derived from the NH band around 336\,nm. 
In \citet{SpiteCH06} it has been shown that the nitrogen abundance based on the Kurucz data, is, on average, 0.4\,dex higher than the nitrogen abundance deduced from the CN band. At that time, since the data of the CN band seemed more robust, a correction of --0.4 dex was applied to the abundance deduced from the NH band. A similar correction was also used in the determination of the nitrogen abundance in GCs \citep{PasquiniEB08}. \\
But more recently \citet{FernandoBH18} presented a new linelist of the $\rm A^{3}\Pi-X^{3}\Sigma^{-}$  electronic transition of NH based on high level calculations. This new linelist seems to correct this effect \citep[see section 3.3 in ][]{FernandoBH18} and has been adopted to compute the N abundance in our stars. The difference between the N abundance measured from the \citet{FernandoBH18} linelist and the Kurucz linelist is 0.50\,dex  in the region 3357-3365\,\AA. However, it can be seen in Fig.\,\ref{Fig:compNHband} 
that the difference is a little smaller around 3364\,\AA~ (0.44\,dex), and a little larger (0.52\,dex) around 3359\,\AA.\\

\begin{figure*}
\begin{center}
\resizebox{8.5cm}{!}
{\includegraphics [clip=true]{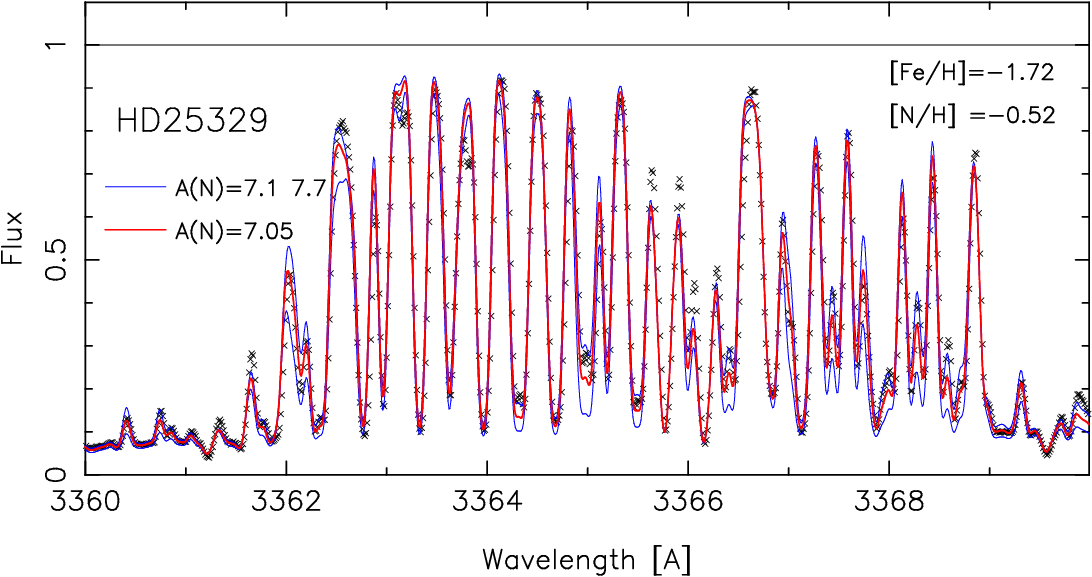}}
\resizebox{8.5cm}{!}
{\includegraphics [clip=true]{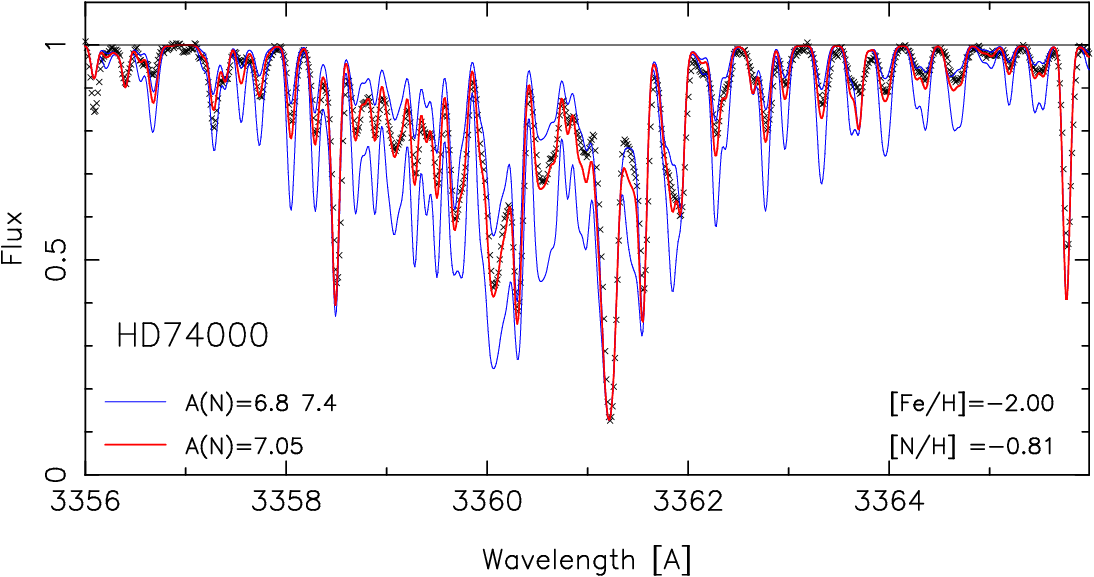}}
\resizebox{8.5cm}{!}
{\includegraphics [clip=true]{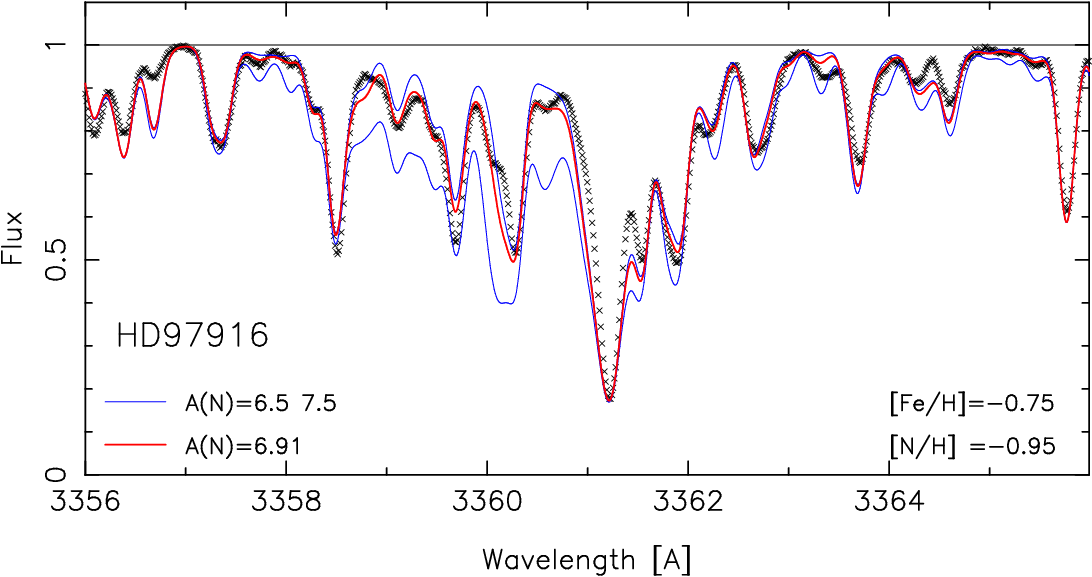}}
\resizebox{8.5cm}{!}
{\includegraphics [clip=true]{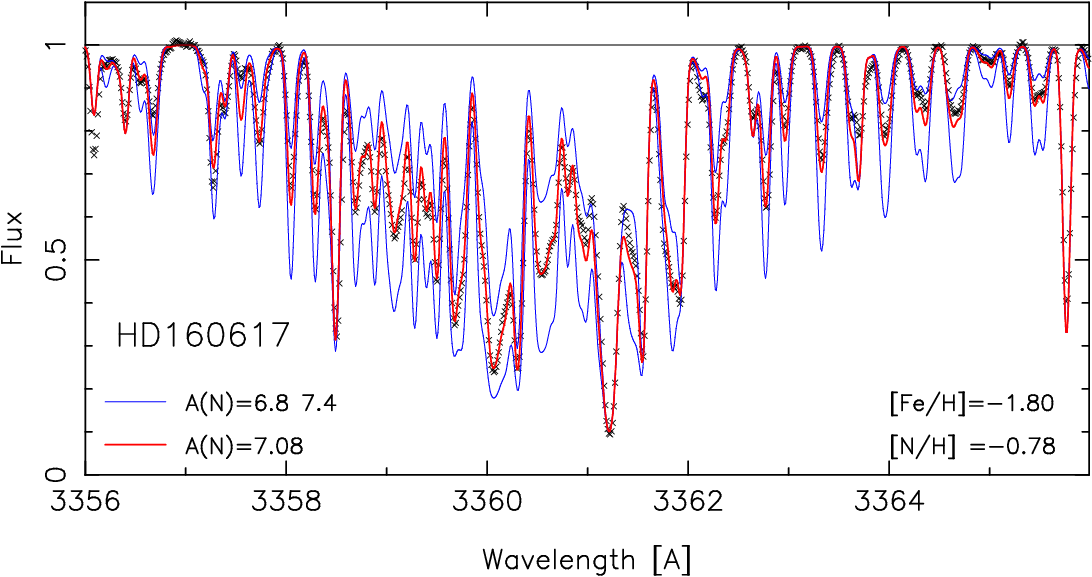}}
\resizebox{8.5cm}{!}
{\includegraphics [clip=true]{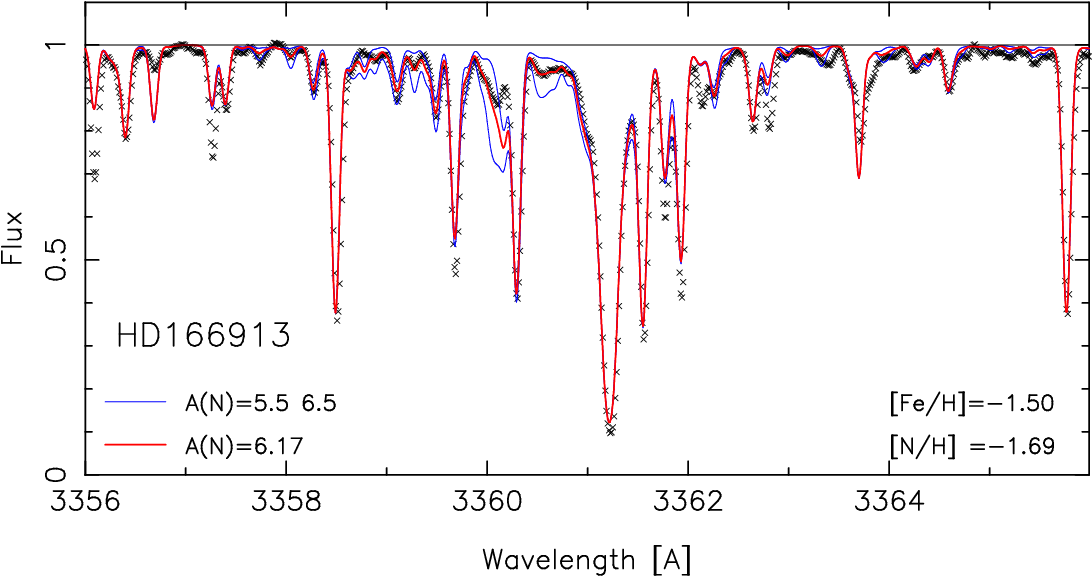}}
\resizebox{8.5cm}{!}
{\includegraphics [clip=true]{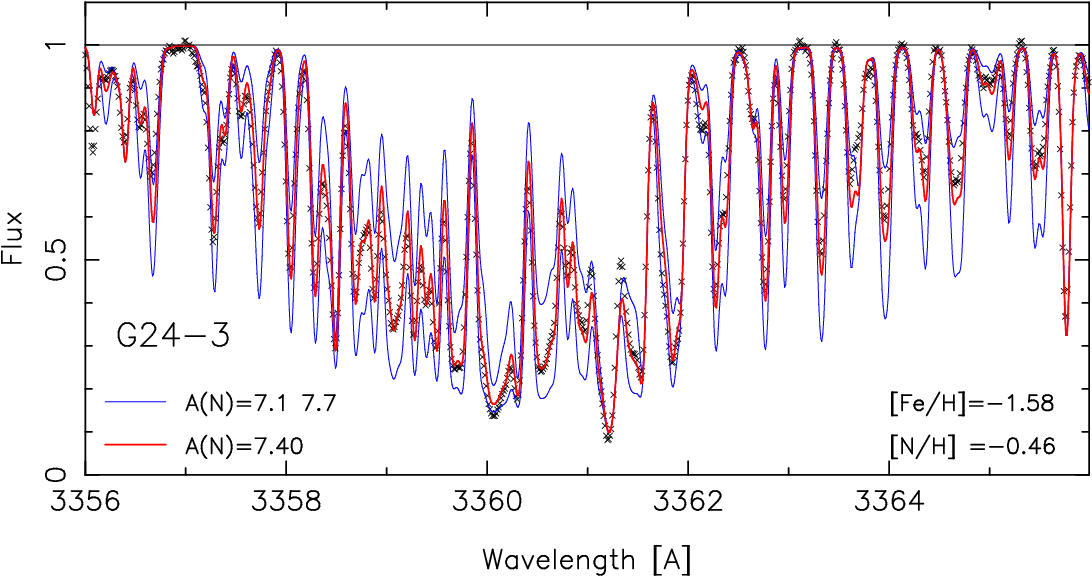}}
\resizebox{8.5cm}{!}
{\includegraphics [clip=true]{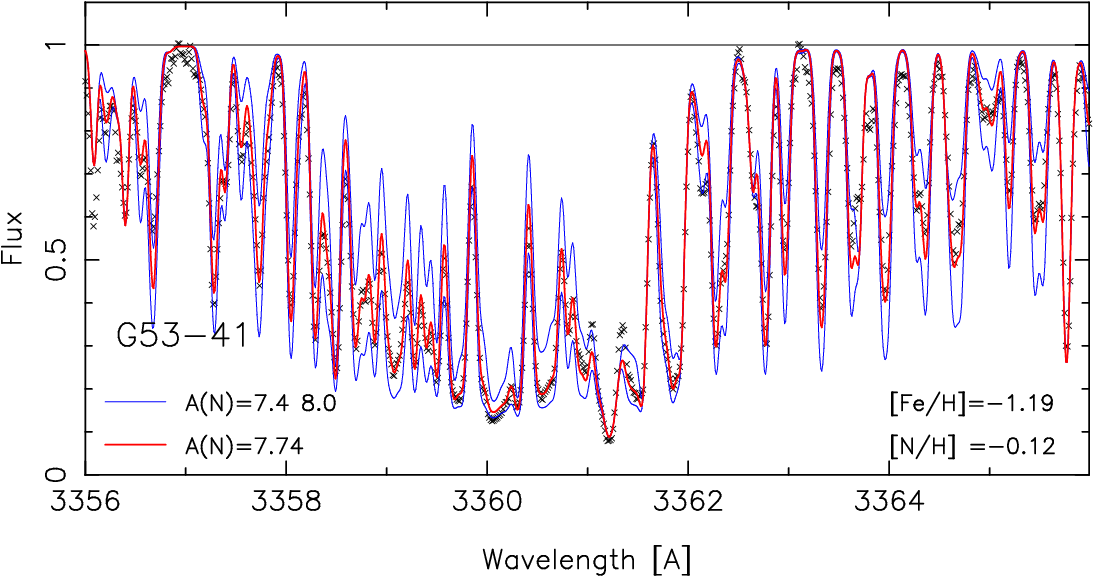}}
\resizebox{8.5cm}{!}
{\includegraphics [clip=true]{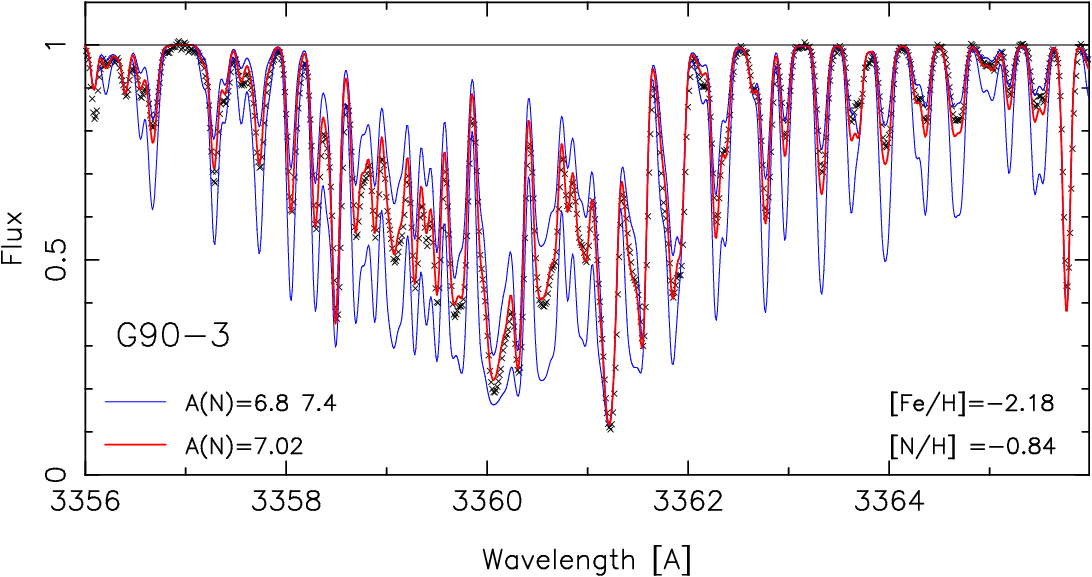}}
\end{center}
\caption[]{Fit of the NH band in our sample of stars known or suspected to be N-rich. The small black crosses are the observed spectrum. The red line represents the best fit of the observed spectrum, and the blue lines synthetic spectra computed with two different values of A(N) indicated on the figure. For HD\,25329, the cooler dwarf of our sample, the region between 3358 and  3362\AA\, is saturated and thus we fitted the observed spectra between 3366 and 3369\AA. The values of [Fe/H] and [N/H] are indicated for each star on the figure.}
\label{Fig:NHband}
\end{figure*}

In Fig.\,\ref{Fig:NHband} we present, for the eight stars suspected to be N-rich, the fit of the observed spectrum with synthetic profiles computed with the NH data of \citet{FernandoBH18}. In this sample two stars HD\,97916 and HD\,166913 with $\rm [N/H]\leq [Fe/H]$ are not N-rich stars.

One of our stars, HD\,160617, has been already extensively studied by \citet{RoedererLawler12} in order to determine the abundance distribution of the r-process elements in this star. They used a very similar model atmosphere, and, as expected, there is a good agreement between their abundance determinations and those given in Table\,\ref{Tab:abundNH}, with one exception: the abundance of nitrogen. While \citet{RoedererLawler12} found A(N)=6.61 we measured A(N)=7.08. The difference is very probably due to the NH line list used by \citet{RoedererLawler12}.

The adopted oxygen abundance has been determined as the mean of the abundance deduced from the  oxygen triplet of \Ou~ (when these lines are not out of the extent of our spectra, see Table \ref{Tab:spectra}), and of the abundance deduced from the OH band around 313\,nm \citep{RichBoesgaard09} or  330\,nm.  
As for the CH band, the parameters of the OH band were directly retrieved from the Bertrand Plez site; they are based on the Kurucz line list.

\subsection{Comparison of the chemical composition of the NH-rich dwarfs and the normal metal-poor dwarf}
To better estimate the peculiarities of the NH-rich stars, we first compared their chemical composition to the composition of normal turnoff metal-poor stars with about the same atmospheric parameters (temperature, gravity and metallicity). We chose a sample of metal turnoff stars studied with the same methods as described in section \ref{Sec:ab1} in the frame of the paper of \citet{PetersonBS20}. The parameters of their models and their chemical abundances are given in Table \ref{Tab:abundNor}. These stars were mainly observed with UVES. For HD\,19445 a complementary spectrum was obtained with the spectrograph NARVAL at the Observatoire du Pic du Midi. For HD\,140283 the abundances given by \citet {SiqueiraAB15} were adopted, we only added the N and O abundances measured from the NH and OH bands on UVES spectra.

\begin{figure}
\begin{center}
\resizebox{8.0cm}{!}
{\includegraphics [clip=true]{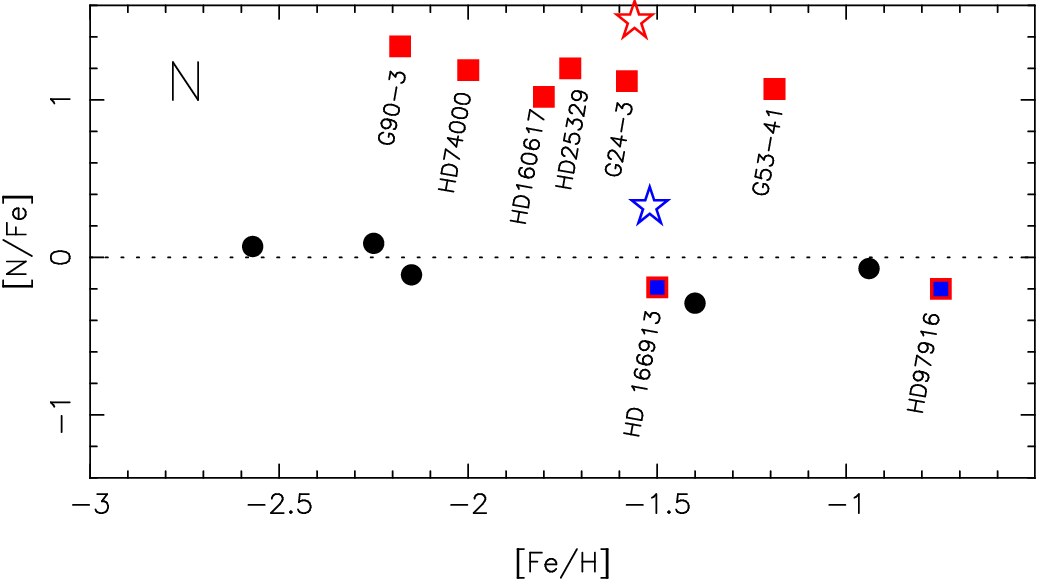}}
\end{center}
\caption{[N/Fe] vs. [Fe/H] for the stars studied in the present paper. Red filled square: N-rich stars. Blue dots surrounded by a red square: stars reported as being N-rich from low resolution spectra and finally found to have a normal nitrogen abundance. Black dots: normal metal-poor turnoff stars.  The blue and red open star symbols represent two dwarf stars in NGC6752 discussed in section \ref{sec:compGC}.
}
\label{Fig:Nab}
\end{figure}

\subsubsection{Nitrogen abundance}
In  Fig.\,\ref{Fig:Nab} we present the [N/Fe] ratio as a function of [Fe/H] in the normal field turnoff stars (black dots) and in the eight stars suspected to be N-rich. 

Among these eight stars, six have indeed a very high abundance of nitrogen with a ratio [N/Fe]>1.0\,dex  (Table \ref{Tab:abundNH}), they are represented by red squares in Fig.\,\ref{Fig:Nab}. But two of them, HD\,97916 and HD\,166913, suspected to be N-rich by \citet{Laird85}, have a normal N abundance with $\rm[N/Fe] \leq 0.0$  (blue dots surrounded by red squares in Fig.\,\ref{Fig:Nab}). In Table\,\ref{Tab:abundNH}, these two stars were put at the end of the list of stars. They will be considered as ``normal'' turnoff stars in the following discussion, and will be represented by blue dots in the subsequent figures.\\

From the Fig.\,\ref{Fig:Nab}  it could be deduced that it exists a gap between the N-rich stars with [N/Fe]>+1.0  and the normal metal-poor stars with [N/Fe] close to zero. In fact, since we have visually chosen turnoff stars with a very strong NH band, there is a strong bias in our selection. Only a large systematic study of the abundance of nitrogen in field metal-poor turnoff stars could allow us to know if all the N-rich stars have a  [N/Fe] ratio close to +1.0 or if there is continuity.

\begin{figure}
\begin{center}
\resizebox{8.0cm}{!}
{\includegraphics [clip=true]{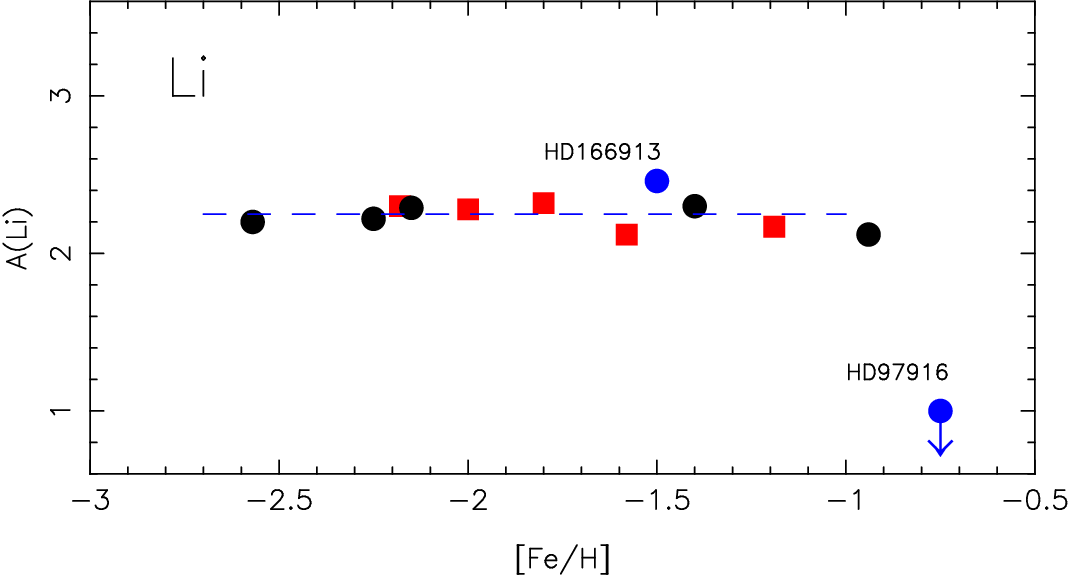}}
\resizebox{8.0cm}{!}
{\includegraphics [clip=true]{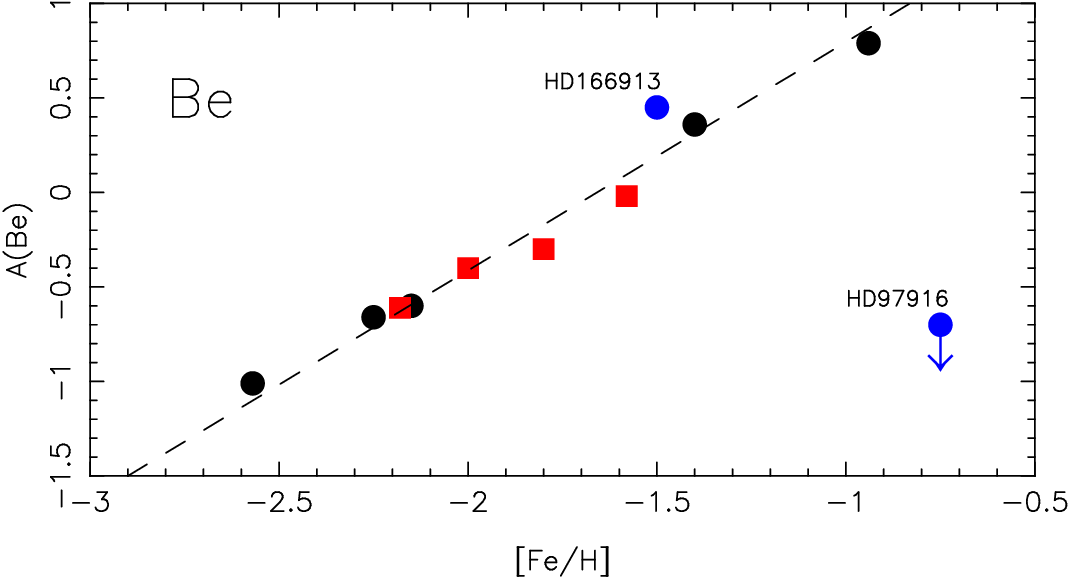}}
\end{center}
\caption[]{A(Li) and A(Be) vs. [Fe/H] in the stars studied in the present paper. Red squares: N-rich stars. Blue dots: stars reported as being N-rich (from low resolution spectra) and finally found to have a normal nitrogen abundance, black dots normal stars.
}
\label{Fig:libe}
\end{figure}

\subsubsection{Abundance of the light elements Li Be}  \label{Sec:liab}
In Fig.\,\ref{Fig:libe}, we compare the lithium and beryllium abundance A(Li) and A(Be)  in  normal metal-poor dwarfs and in the NH-rich dwarfs.  
With the temperature scale adopted, based on the Gaia photometry and the PARSEC isochrones, no clear systematic difference between the N-rich stars and the normal turnoff stars is visible. This result is, on a more extended sample, in agreement with the result of  \citet{SpiteSpite86}. We note that HD\,25329 has no detectable Li, but  the temperature of this star is only 4870K, thus, the convective zone in the atmosphere is deep and Li is brought to hot layers where it is little by little destroyed. This star is not plotted in Fig.\,\ref{Fig:libe}.\\
-- In addition we note that in the ``normal'' star HD 97916, Li is not detected and it is very  Be-poor. We derived A(Be)<-0.7 and  \citet{Boesgaard07} who measured A(Be)<-1.3 considers that it is a blue straggler. The fact that we found a very young age for this star (section \ref{sec:Par}) reinforces this interpretation.\\
-- Moreover we note that the ``normal'' star, HD\,166913 ([Fe/H)=--1.5)  has a high Li abundance: A(Li)=2.46 (see Fig.\,\ref{Fig:libe}, upper panel). The Be abundance is also rather high in this star (Fig.\,\ref{Fig:libe}, lower panel). The ratio \lisisept~ estimated from the UVES spectrum could be close to 0.1, but the resolution of the spectrum ($R=40000$ in the region of the Li feature) is not sufficient to be confident in this result. An EXPRESSO spectrum of this star ($R=150000$) has been recently obtained and is under study. If confirmed, these anomalies (rather high values of Li and Be and presence of $\rm^{6}Li$) could be explained by the engulfment of a planet.

\begin{figure}
\begin{center}
\resizebox{8.0cm}{!}
{\includegraphics [clip=true]{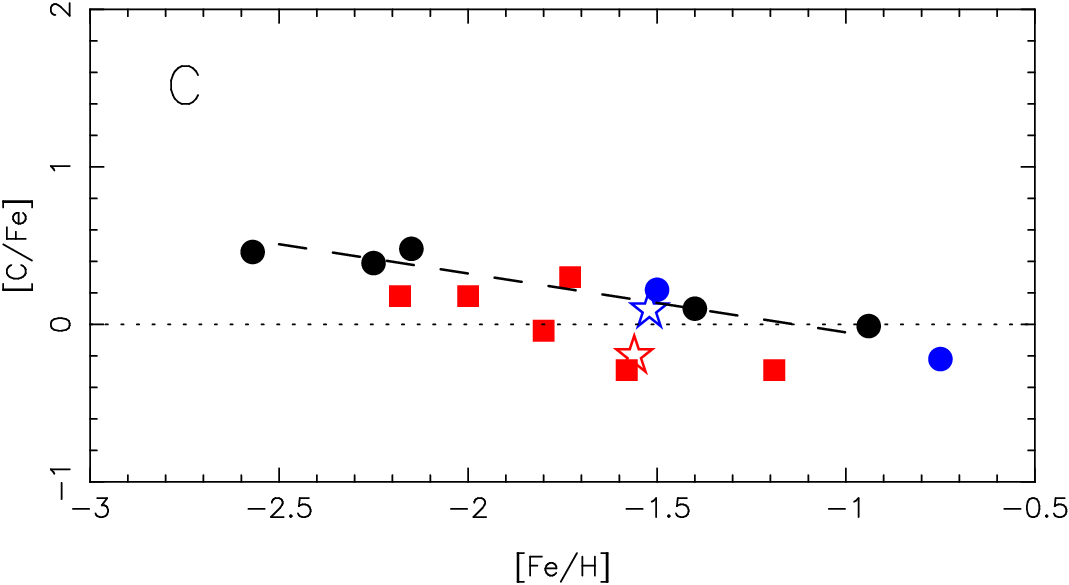}}
\resizebox{8.0cm}{!}
{\includegraphics [clip=true]{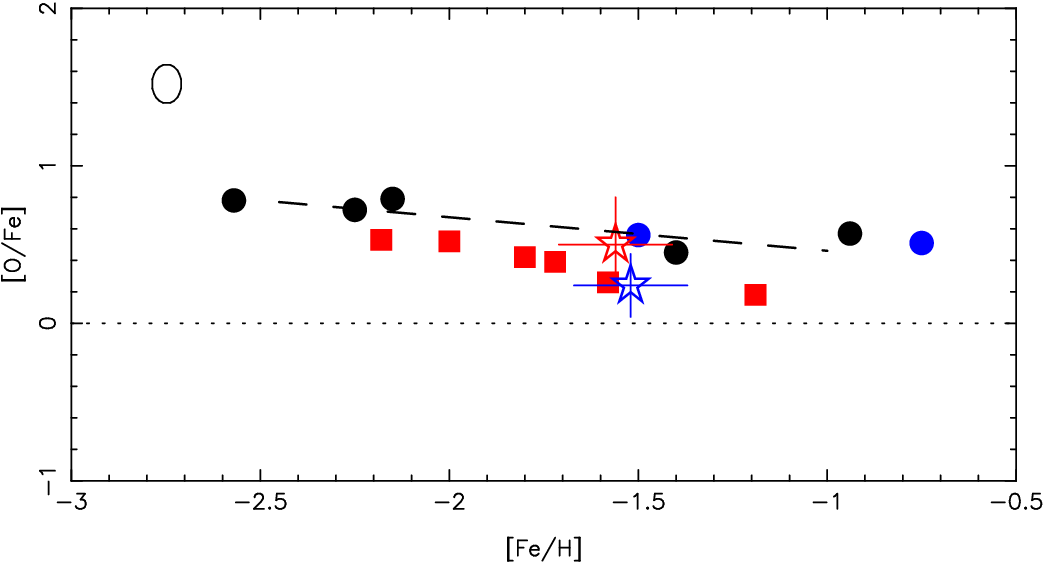}}
\end{center}
\caption[]{[C/Fe] and [O/Fe] vs. [Fe/H] in the stars studied in the present paper. The symbols are the same as in Fig.\,\ref{Fig:libe}. The blue star symbol represent NGC6752-4428 a first generation dwarf in NGC6752, and the red star symbol NGC6752-200613 a second generation dwarf (see section \ref{sec:compGC} where these stars will be discussed).
}
\label{Fig:abCO}
\end{figure}

\begin{figure}
\begin{center}
\resizebox{8.0cm}{!}
{\includegraphics [clip=true]{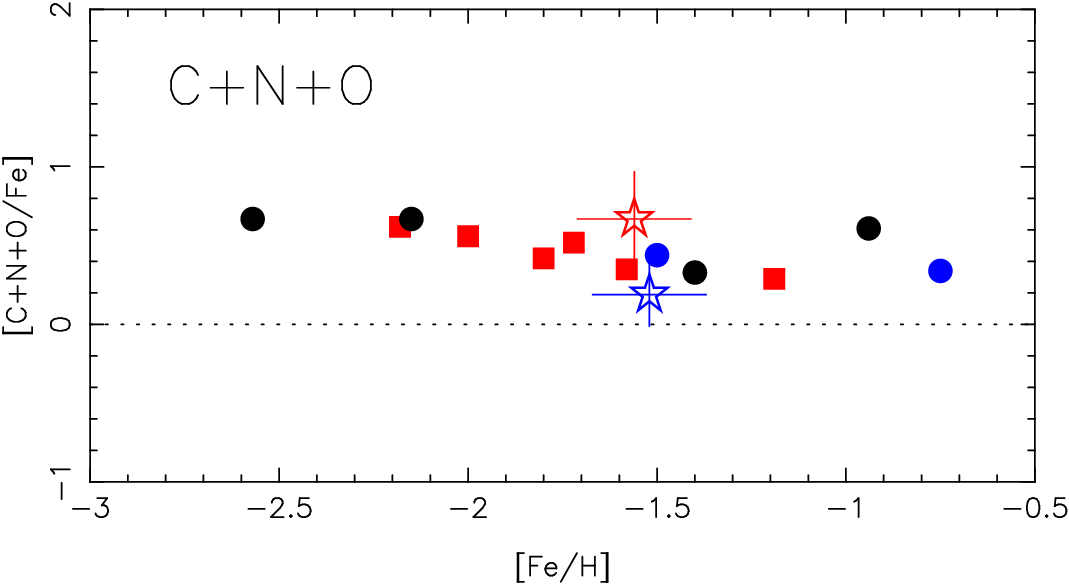}}
\end{center}
\caption[]{[(C+N+O)/Fe] in our sample of stars. The symbols are the same as in Fig.\,\ref{Fig:abCO}. The [(C+N+O)/Fe] ratio is about the same in normal and in N-rich stars.
 }
\label{Fig:cno}
\end{figure}

\subsubsection{Abundance of the elements from C to O}
In the interval $\rm -2.5<[Fe/H]-0.5$, the ratios [C/Fe] and [O/Fe] increase when the metallicity decreases  (see Fig.\,\ref{Fig:abCO}), but the abundances of C and O in N-rich stars are systematically lower by about 0.3\,dex than the abundances in the comparison stars (see Fig.\,(Fig.\,\ref{Fig:cno}) 
the N-rich stars have practically the same [(C+N+O)/Fe] value as the normal stars. This suggests that the nitrogen in N-rich stars was largely formed at the expense of C and O.

\begin{figure}
\begin{center}
\resizebox{8.0cm}{!}
{\includegraphics [clip=true]{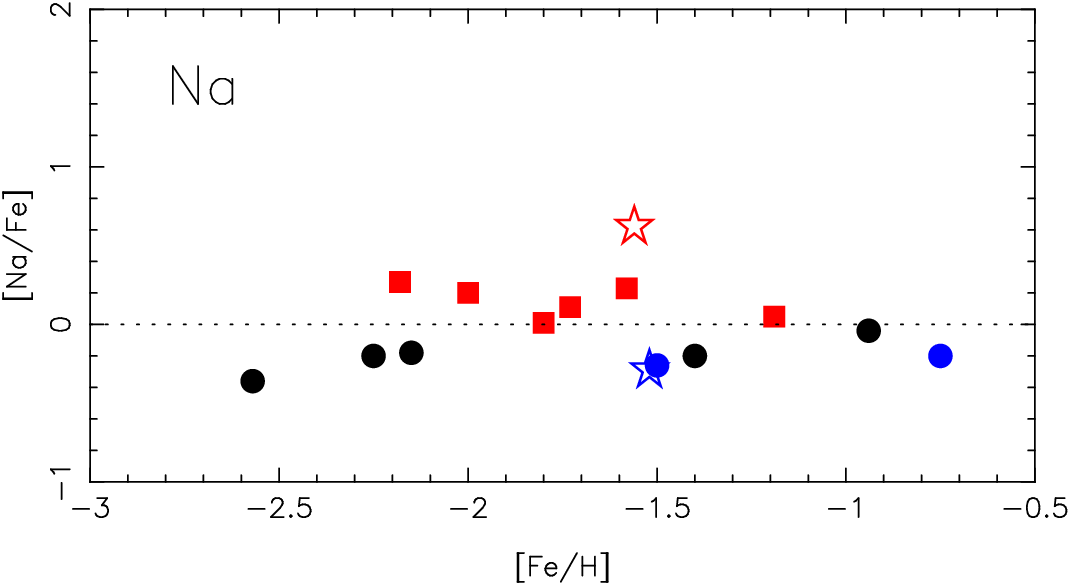}}
\resizebox{8.0cm}{!}
{\includegraphics [clip=true]{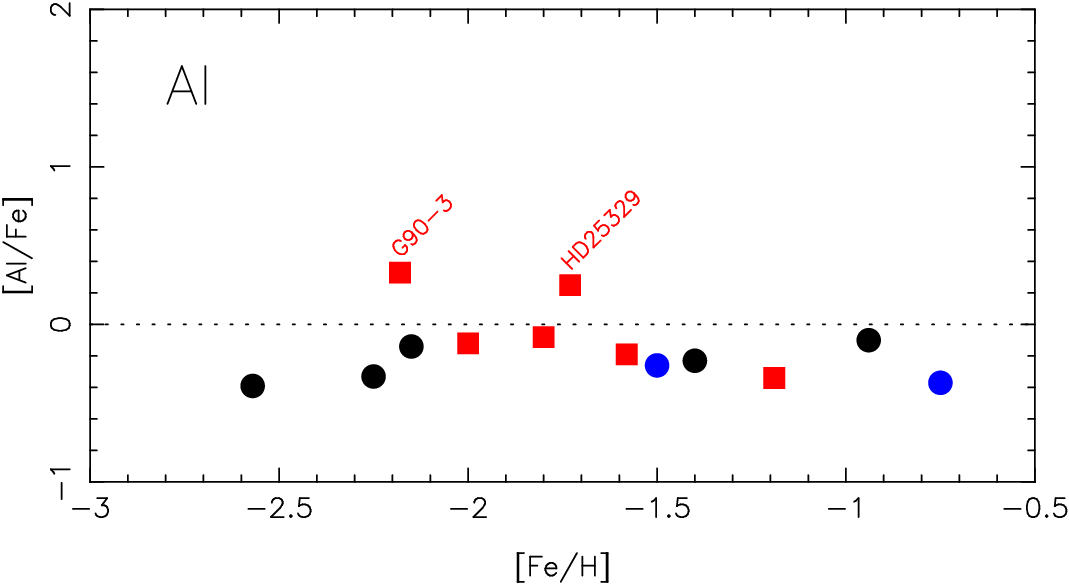}}
\end{center}
\caption[]{[Na/Fe] and [Al/Fe] vs. [Fe/H] in the stars studied in the present paper. The symbols are the same as in Fig.\,\ref{Fig:abCO}}
\label{Fig:abNaAl}
\end{figure}

\begin{figure*}
\begin{center}
\resizebox{6.0cm}{!}
{\includegraphics [clip=true]{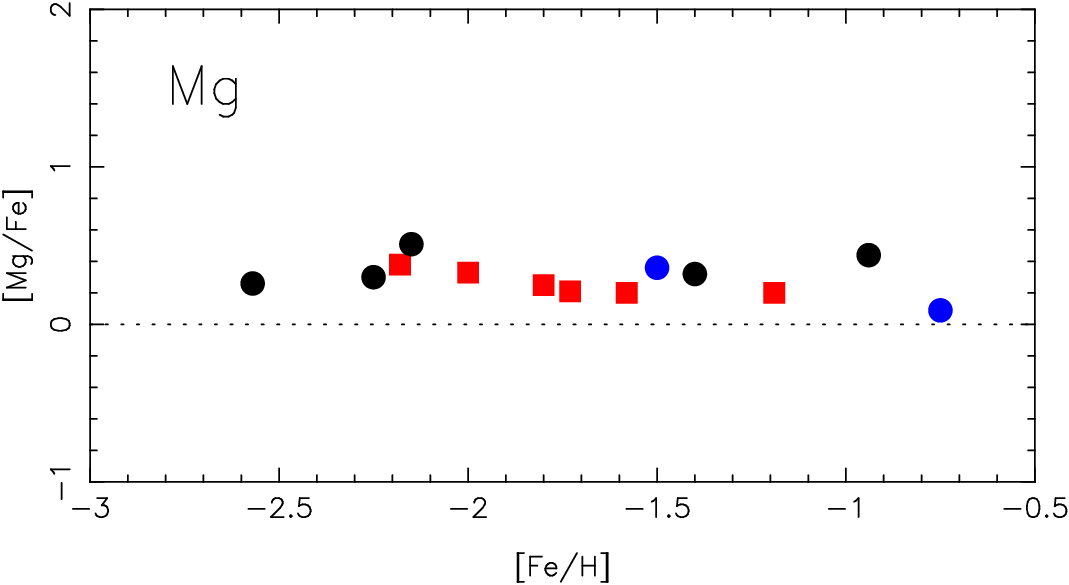}}
\resizebox{6.0cm}{!}
{\includegraphics [clip=true]{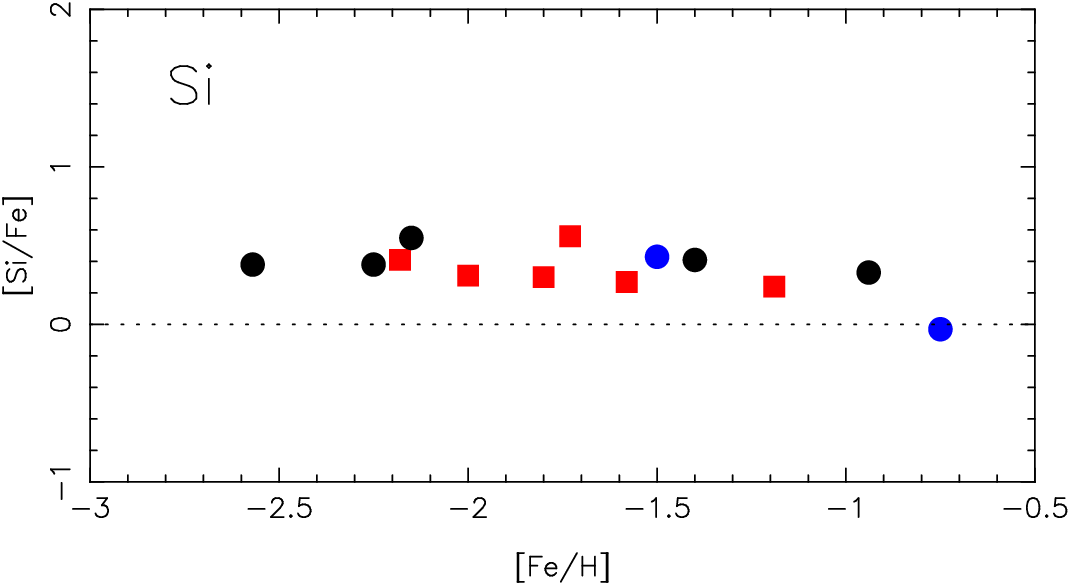}}
\resizebox{6.0cm}{!}
{\includegraphics [clip=true]{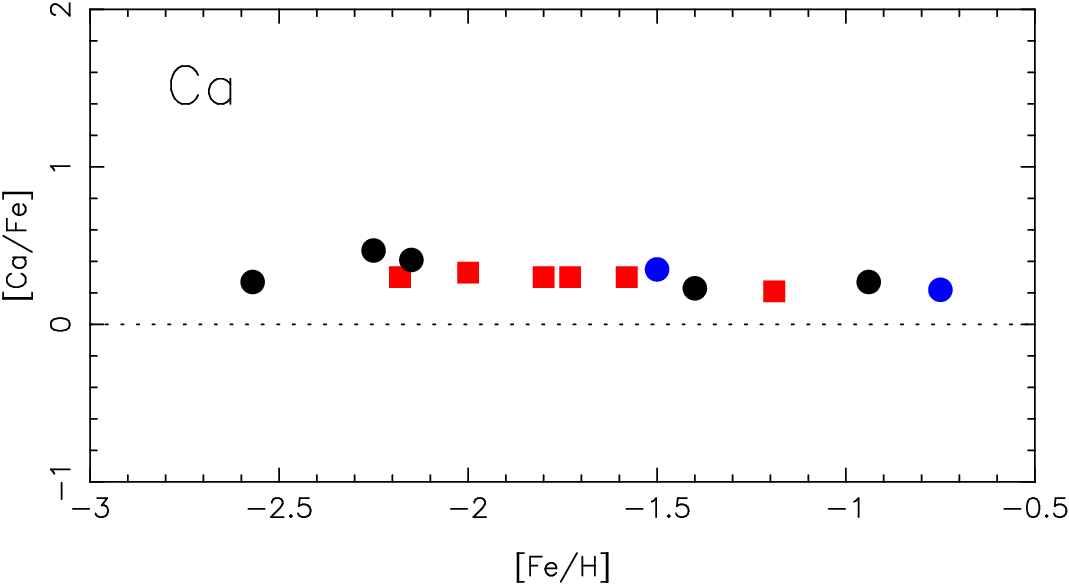}}
\resizebox{6.0cm}{!}
{\includegraphics [clip=true]{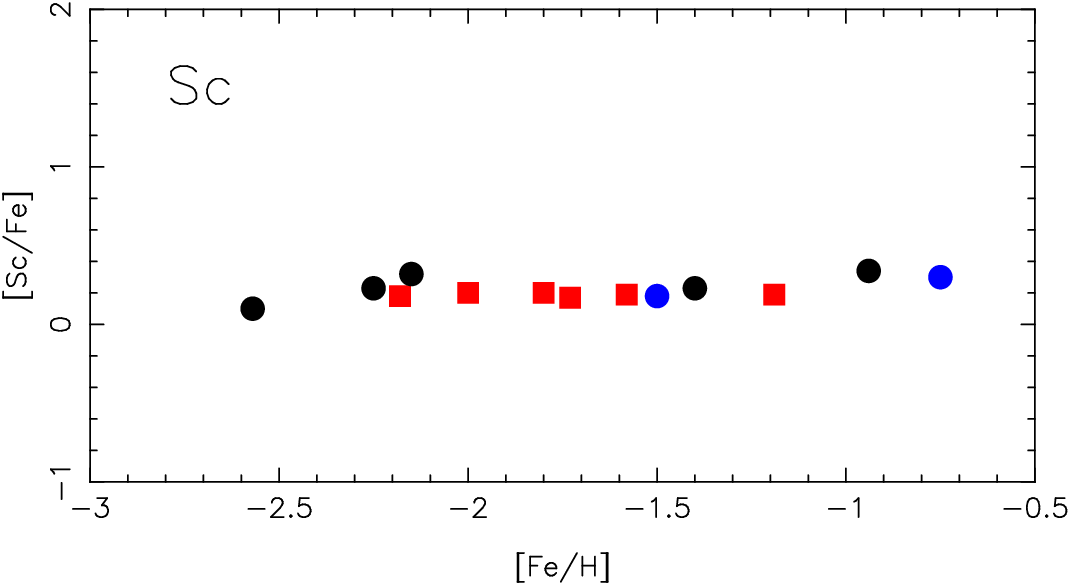}}
\resizebox{6.0cm}{!}
{\includegraphics [clip=true]{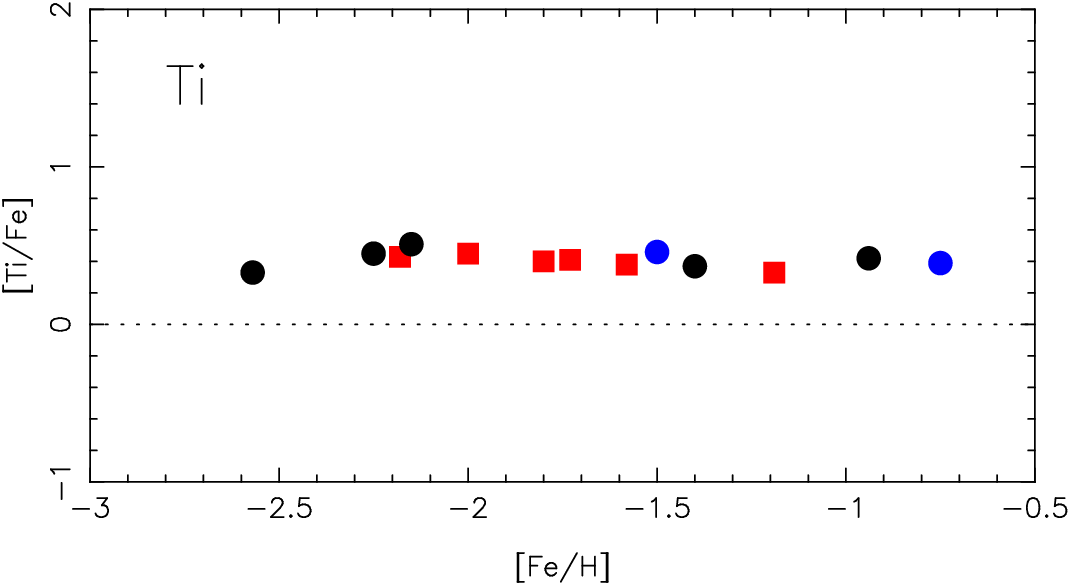}}
\resizebox{6.0cm}{!}
{\includegraphics [clip=true]{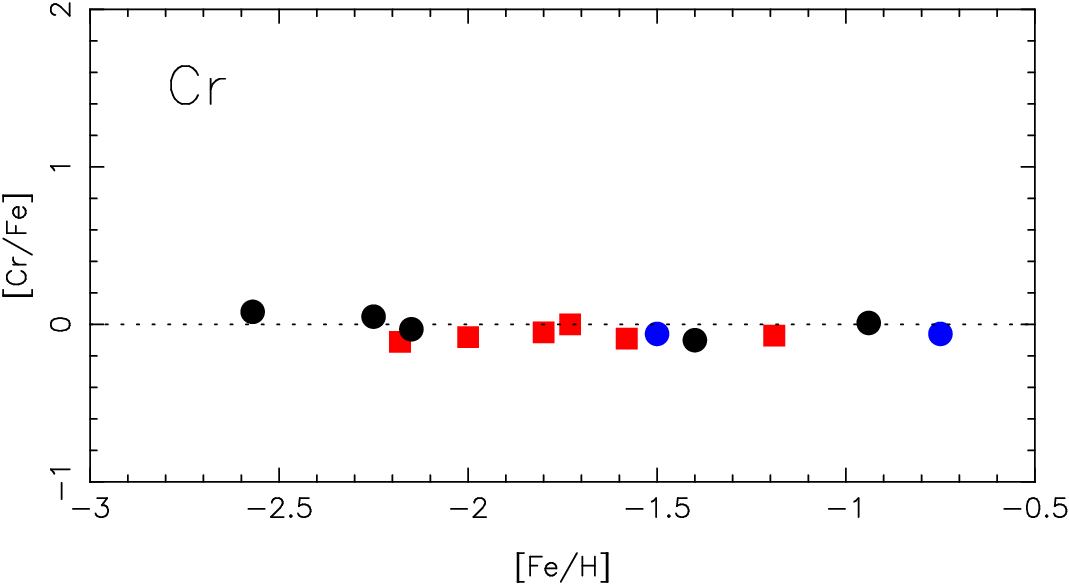}}
\resizebox{6.0cm}{!}
{\includegraphics [clip=true]{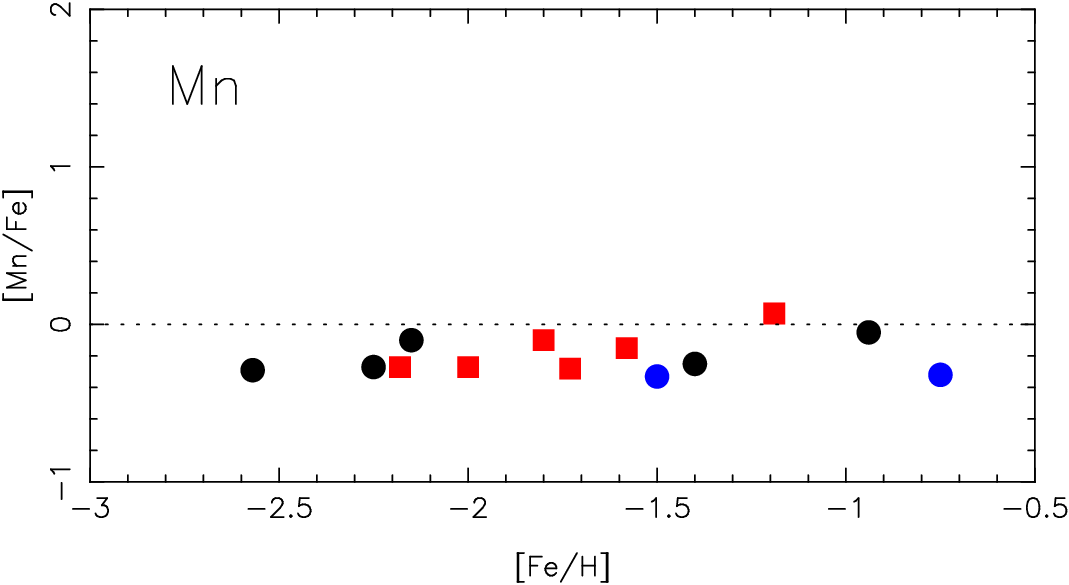}}
\resizebox{6.0cm}{!}
{\includegraphics [clip=true]{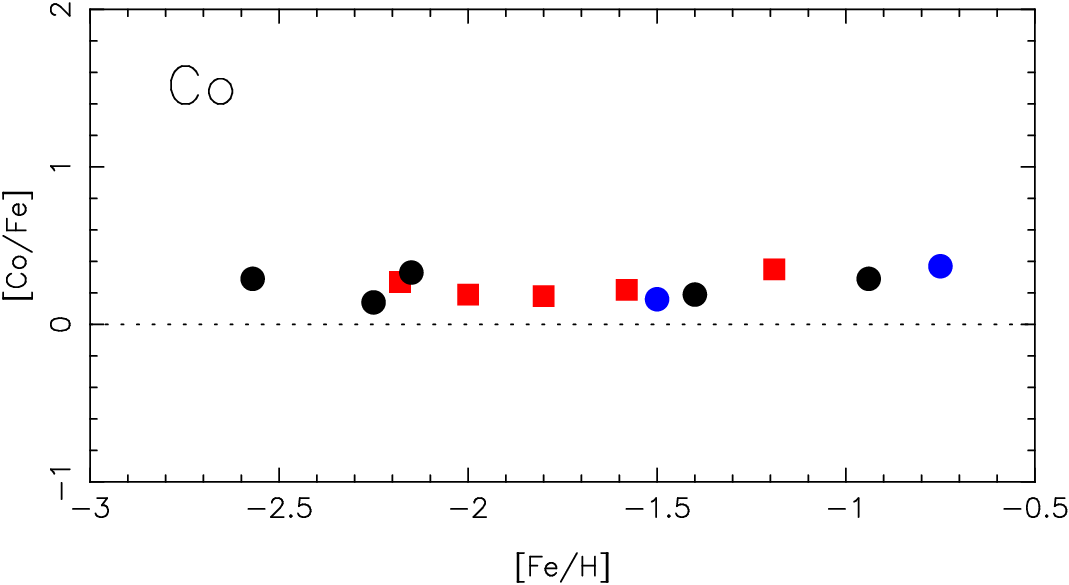}}
\resizebox{6.0cm}{!}
{\includegraphics [clip=true]{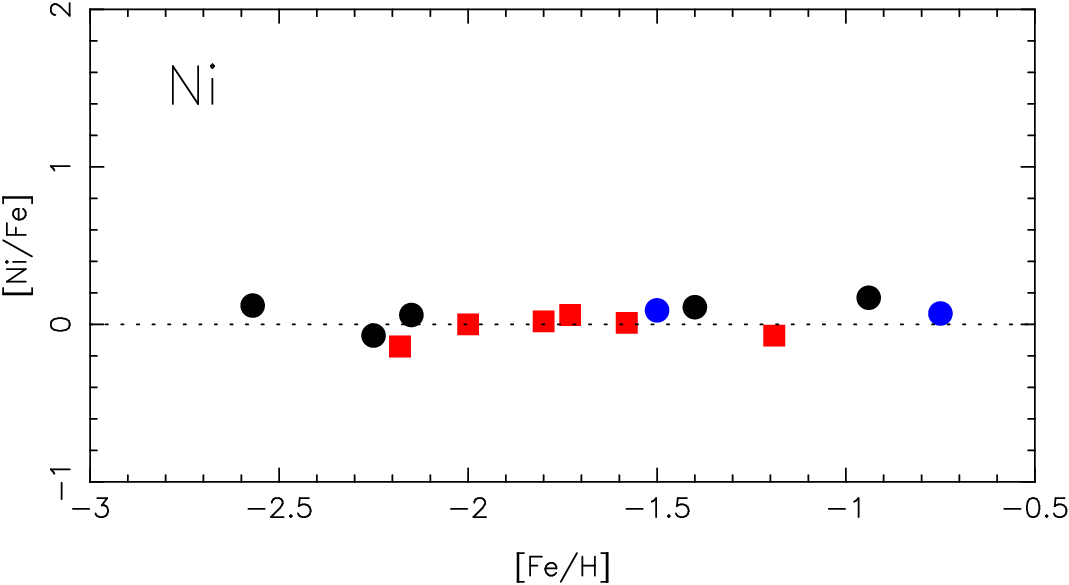}}
\resizebox{6.0cm}{!}
{\includegraphics [clip=true]{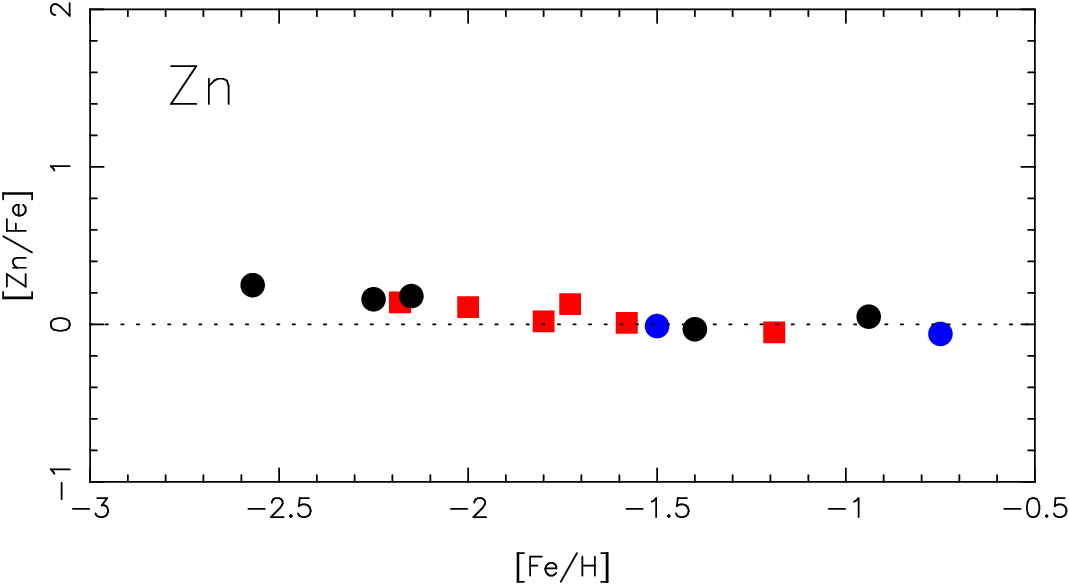}}
\resizebox{6.0cm}{!}
{\includegraphics [clip=true]{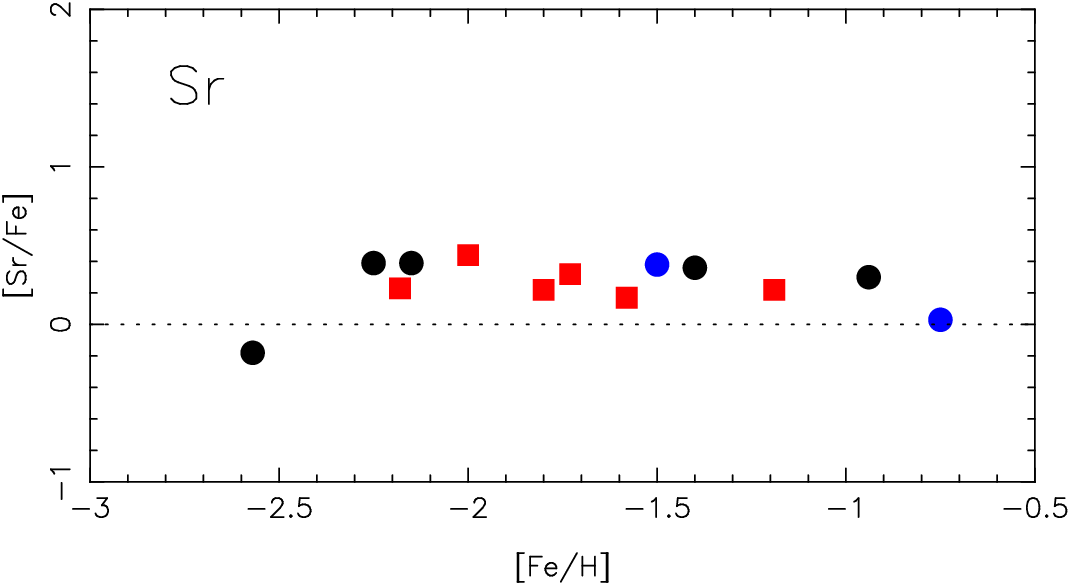}}
\resizebox{6.0cm}{!}
{\includegraphics [clip=true]{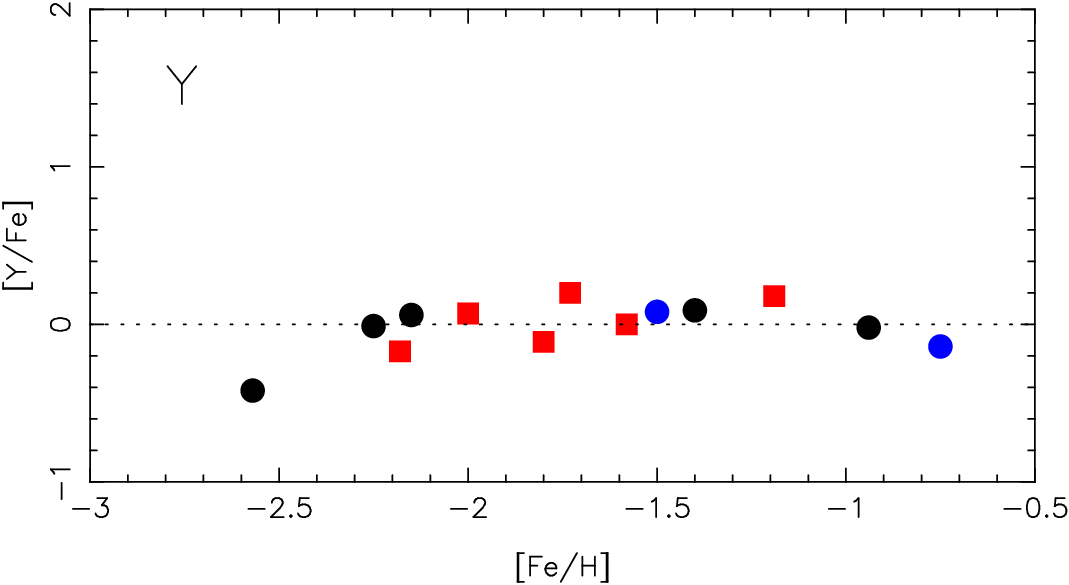}}
\resizebox{6.0cm}{!}
{\includegraphics [clip=true]{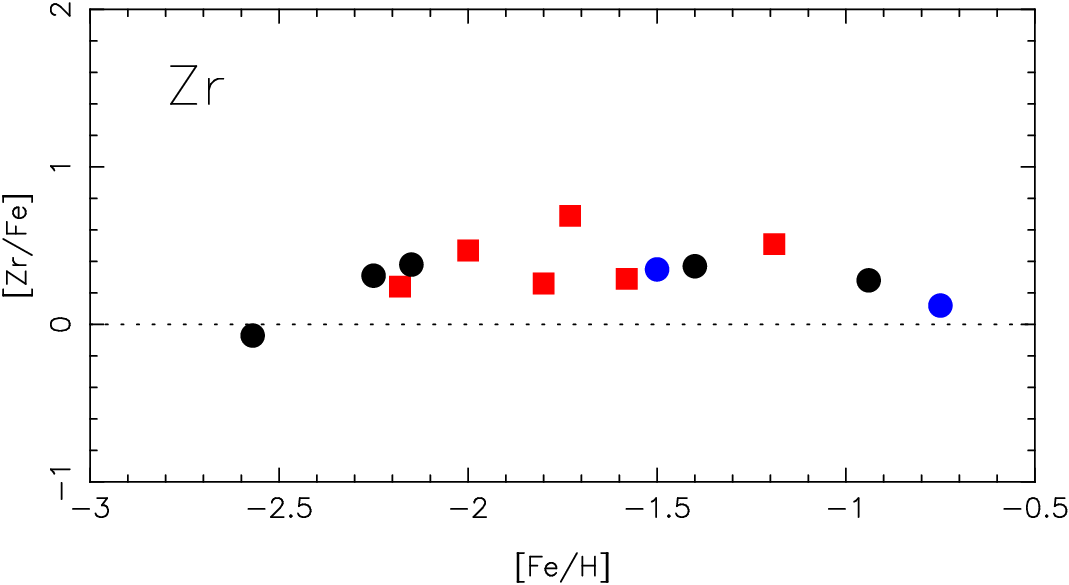}}
\resizebox{6.0cm}{!}
{\includegraphics [clip=true]{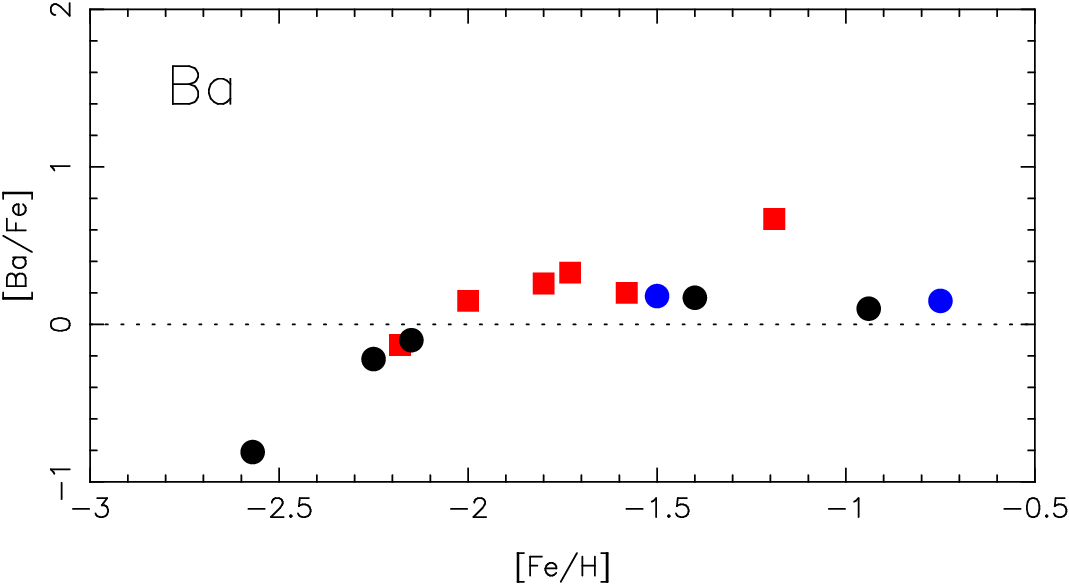}}
\resizebox{6.0cm}{!}
{\includegraphics [clip=true]{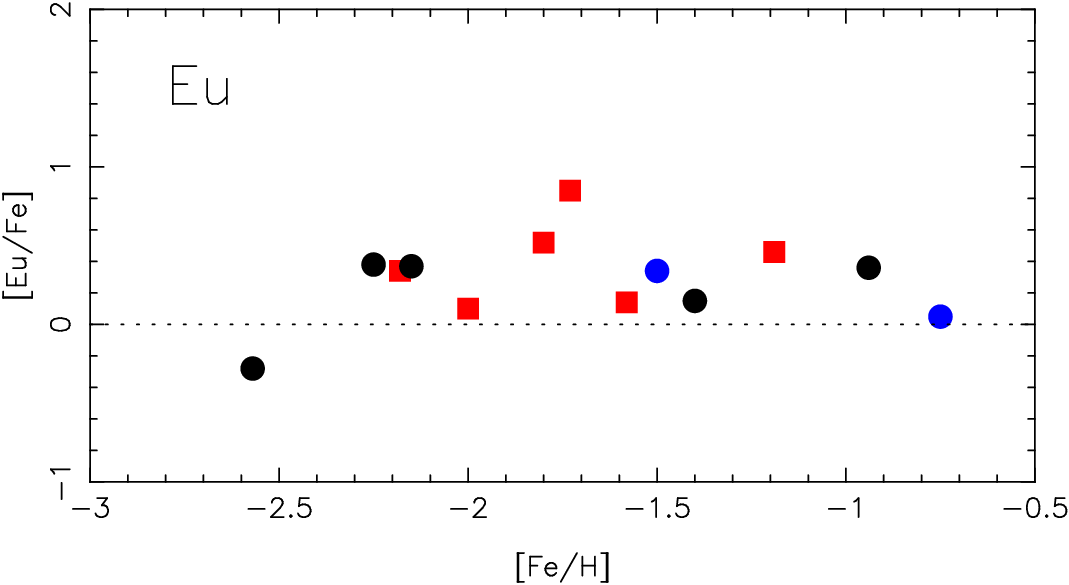}}
\end{center}
\caption[]{[X/Fe] vs. [Fe/H]  for the $\alpha$ elements (Mg to Ti), the iron peak elements (Cr to Zn) and the neutron capture elements (Sr to Eu). The symbols are the same as in Fig.\,\ref{Fig:abCO}.}
\label{Fig:abMg-Eu}
\end{figure*}

\subsubsection{Abundance of the odd elements Na, Al} 
In Fig.\,\ref{Fig:abNaAl} we compare the abundance of Na and Al in N-rich stars and in normal metal-poor stars. We observe that [Na/Fe] is systematically higher in the N-rich stars. It seems that there is also an overabundance of Al in some of these stars, and it reaches almost 0.4\,dex in G90-3.\\
 \citet{ZhaoMY16} had already remarked the high abundance of Na in HD\,74000 and in G90-03  and the high Al abundance in G90-03.

\subsubsection{$\alpha$ and iron-peak elements}
In nitrogen-rich stars, as in ``normal'' metal-poor stars, the $\alpha$ ~elements Mg, Si, Ca, Ti are enhanced  by about 0.4\,dex (Fig.\,\ref{Fig:abMg-Eu}). The iron peak elements behave also as in normal stars: Cr and Ni have solar [X/Fe] ratios as well as Zn, [Co/Fe] is close to +0.2\,dex and [Mn/Fe] is subsolar ($\rm \simeq -0.2\,dex$). 

\subsubsection{Heavy elements}
The elements heavier than Zn are formed by addition of neutrons on iron peak elements. 
There are two basic mechanisms able to capture neutrons on existing seed nuclei: a first one, rapid relative to the average $\beta$ decay (``r'' process), happens in core collapse supernovae or during the merging of neutron stars;  another one, much slower,  (``s'' process) happens in thermally pulsing asymptotic giant branch (AGB) stars. \\
In metal-poor stars the formation by the r-process dominates and the influence of the ``s'' process appears only for  $\rm [Fe/H] \geq -1.4$ dex \citep{RoedererCK10}.
For these heavy elements, the ratio [X/Fe] is more scattered than it is for the $\alpha$ or the iron-peak elements \citep[see for example][]{FrancoisDH07},  this large scatter is observed in Fig.\,\ref{Fig:abMg-Eu} in particular for Ba and Eu. It is well known that some metal-poor stars are rich in Eu and Ba (r-rich stars) and other are Eu and Ba-poor (r-poor stars). HD\,140283, the most metal-poor ``normal'' star in Fig.\,\ref{Fig:abMg-Eu}, with $\rm [Fe/H] \simeq -2.6$,  is a classical ``r-poor'' metal-poor star with a very low abundance of Ba and Eu. At the same metallicity ($\rm[Fe/H] \simeq -2.6$) many ``normal'' stars have a higher [Ba/Fe] ratio  \citep[see Fig.\,10 in][]{FrancoisDH07}. 

\begin{figure}
\begin{center}
\resizebox{8.0cm}{!}
{\includegraphics [clip=true]{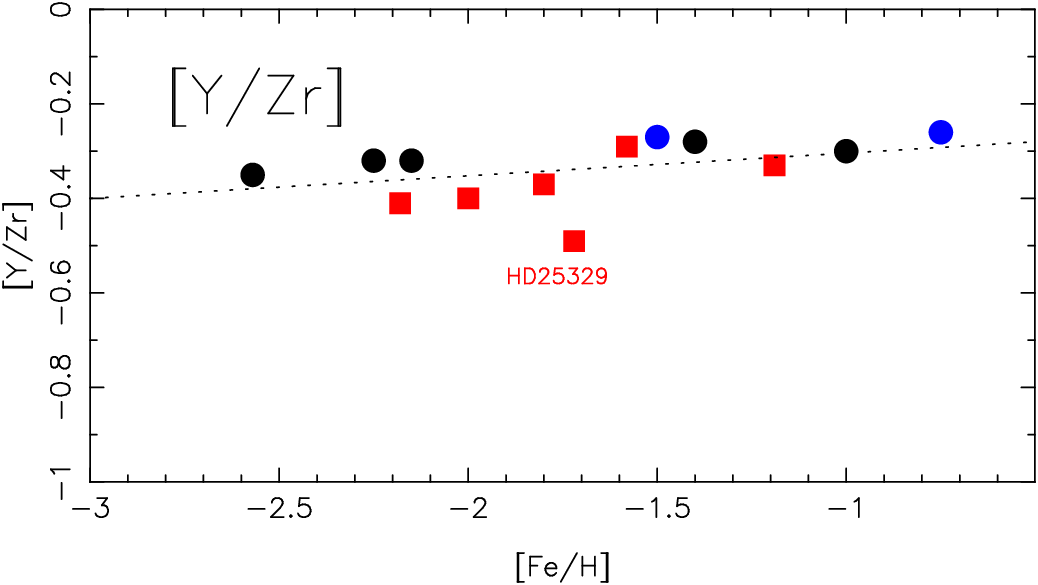}}
\resizebox{8.0cm}{!}
{\includegraphics [clip=true]{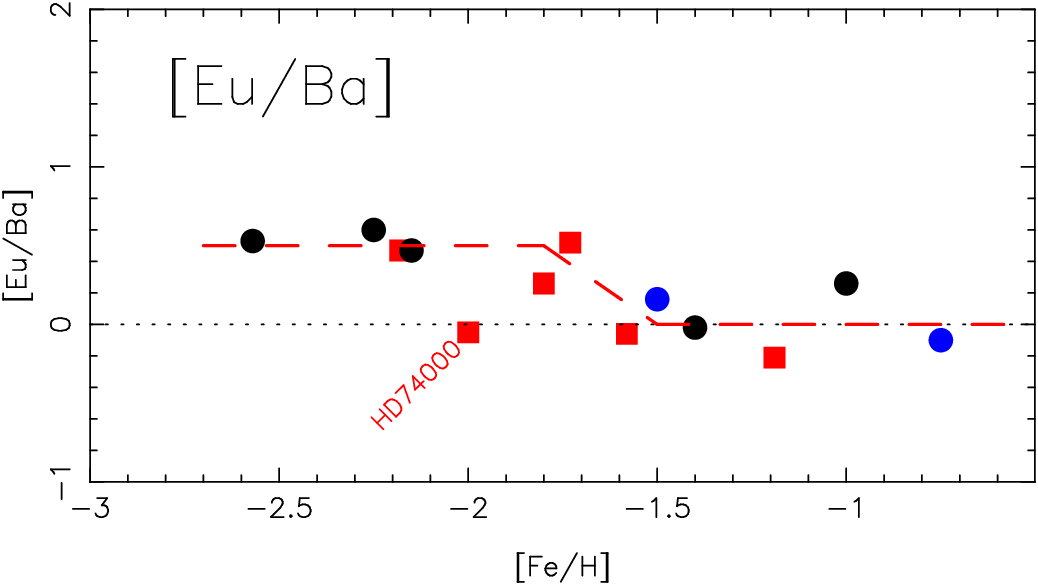}}
\end{center}
\caption[]{ [Zr/Y] and [Eu/Ba] vs. [Fe/H] in the stars studied in the present paper. The symbols are the same as in Fig.\,\ref{Fig:abCO}}
\label{abNHrich-YZrEuBa}
\end{figure}

In Fig.\,\ref{abNHrich-YZrEuBa} (upper panel), we have plotted [Zr/Y] as a function of [Fe/H]. There is a rather good correlation of the abundance of Zr and Y in the normal stars of our sample, but there is some indication of a slightly lower ratio of  [Zr/Y] in the N-rich stars. But this difference has to be confirmed by an homogeneous study of larger samples.

In the most metal-poor stars it has been shown, that  the ratio  [Eu/Ba], is  close to 0.5 dex  \citep[see e.g.][]{FrancoisDH07,Spite-mf14,MashonkinaCB10}. This is also observed in the most metal-poor stars of our sample (Fig.\,\ref{abNHrich-YZrEuBa}, lower panel) with the exception of the N-rich star HD\,74000 which has a solar  [Eu/Ba] abundance ratio, despite its low metallicity ([Fe/H]=--2). This will be discussed later in section \ref{sec:heavyGC}.\\
When [Fe/H]>--1.5, [Eu/Ba] tends to be solar, as expected following \citet{RoedererCK10}.

\section{Comparison of the abundances in the N-rich stars and in the second generation stars of the GCs} \label{sec:compGC}

It is now generally accepted that, inside a GC, first and second generation stars cohabit \citep[see for exemple][]{PrantzosCI07,CharbonnelCK14}. The second generation stars were formed from an intracluster gas polluted by H processed material ejected by a first generation of massive stars. Different objects have been suggested as polluters: mainly fast rotating massive stars (FRMS) with a mass larger than $25 M_{\odot}$,  massive asymptotic giant branch stars with $6<M<11M_{\odot}$, and also massive binaries. The nature of the polluter depends on the initial mass of the GC. Compared to the first generation of stars, the second generation is poor in Li, C and O  but rich in N, Na \citep{CarrettaGL05,PasquiniBM05,PasquiniBR07,PasquiniEB08,LindPC09} and sometimes in Al \citep{MezarosMG20,SchiappacasseLR22}. \\

If the six N-rich stars here studied were born in GCs they certainly do not come from the same cluster and since the chemical anomalies depend on the mass of the cluster, we do not expect a perfect agreement between our sample of stars and the relations observed in a given GC. These field N-rich stars have different metallicities (from --1.2\,dex to --2.2\,dex), but they have almost the same value of [N/Fe] (i.e. $\rm 1.02 \leq [N/Fe] \leq 1.34$) and thus any correlation between [N/Fe] and the other elements cannot be firmly established. However it is interesting to check wether the values observed for Li, C, O, Na, Al in these stars, are comparable to the values in the GCs.

\begin{figure}
\begin{center}
\resizebox{8.0cm}{!}
{\includegraphics [clip=true]{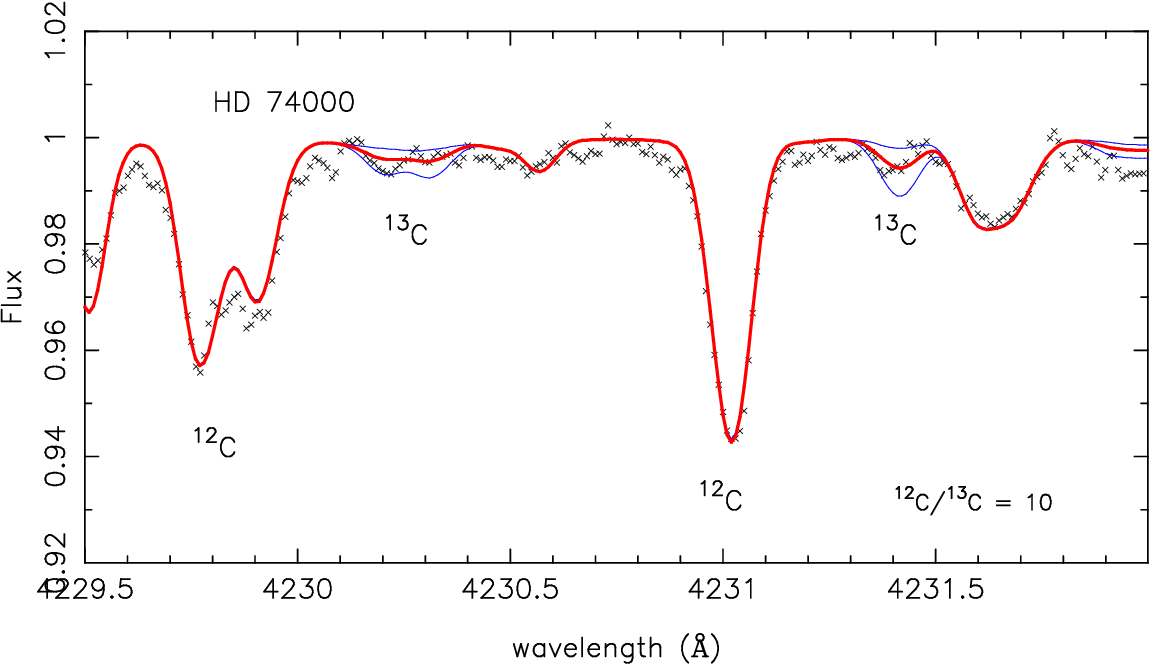}}
\resizebox{8.0cm}{!}
{\includegraphics [clip=true]{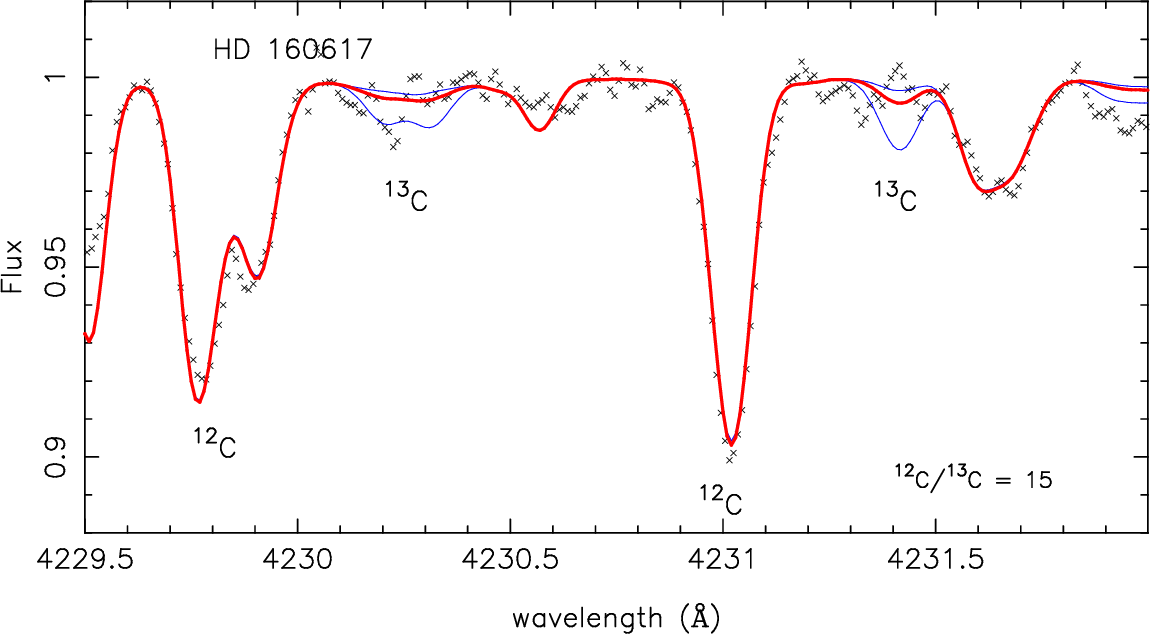}}
\resizebox{8.0cm}{!}
{\includegraphics [clip=true]{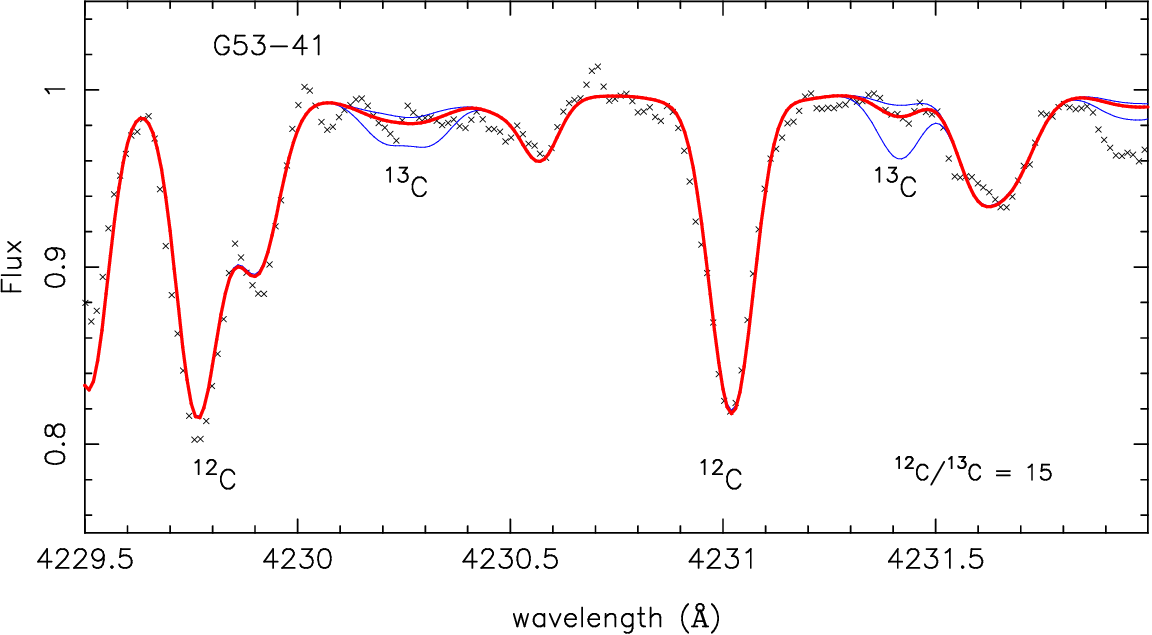}}
\end{center}
\caption[]{\Cdt in three of our sample of stars. The computations have been done for \Cdt=30 \Cdt=5 (blue lines) and for the best fit (red line):  \Cdt=10 for HD 74000, and \Cdt=15 for HD\,160617 and G53-41.}
\label{Fig:cdt}
\end{figure}

\begin{figure}
\begin{center}
\resizebox{7.0cm}{!}
{\includegraphics [clip=true]{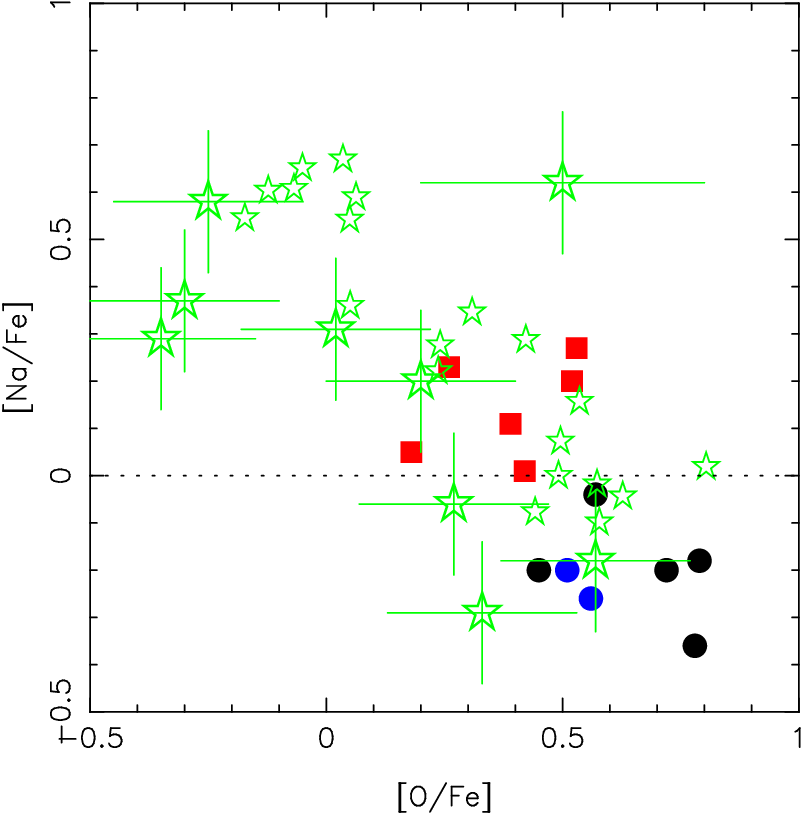}}
\end{center}
\caption[]{The position of our stars in the [Na/Fe] vs. [O/Fe] diagram is compared to the stars of the GC NGC6752. The symbols are the same as in Fig.\,\ref{Fig:abCO}, the giants in NGC6752 are represented by small star symbols, and the dwarfs by larger star symbols. The error is larger in the case of dwarfs since they are fainter and an approximative estimation of this error has been indicated on the figure.}
\label{Fig:NaO}
\end{figure}

\subsection{Comparison of the C, N and O abundances}
We would like to compare the C, N and O abundance in our N-rich stars with these abundances in a second generation dwarfs of a GC. \\
The nitrogen abundance has been measured from the NH band in two dwarf stars of NGC6752: NGC6752-4428 and NGC6752-200613  by \citet{PasquiniEB08}. \citet{CarrettaGL05} measured the N abundance in nine NGC6752 dwarfs from the CN band including the two stars observed by \citet{PasquiniEB08}. There is a rather good agreement for the nitrogen abundance in NGC6752-200613  ([N/Fe]=+1.7 and +1.5) but a clear disagreement for NGC6752-4428 ([N/Fe]=+1.1 and +0.3). Since in metal-poor dwarfs, it is easier to measure the nitrogen  abundance from the NH band we adopted the \citet{PasquiniEB08} values and,  in a first step, we did not take into account the nitrogen abundances computed by \citet{CarrettaGL05}.

The star NGC6752-200613 is strongly enriched in nitrogen ([N/Fe]= 1.5) and \citet{CarrettaGL05} measured the carbon abundance ([C/Fe]=--0.2\,dex), but they could not measure the oxygen abundance from the red oxygen triplet. 
We tried to measure the oxygen abundance from the OH band in the near UV spectrum used by \citet{PasquiniBR07} to measure Be in this star. We obtained $\rm [O/Fe]=+0.5 \pm 0.3$. (The uncertainty is large due to the low S/N in this region of the spectrum.) We can thus compare the C, N, O abundances in NGC6752-200613 and in the stars of our sample. This is done in Fig. \ref{Fig:Nab}, \ref{Fig:abCO}, and \ref{Fig:cno}. NGC6752-200613 is more N-rich than the stars of our sample of N-rich field stars but the general behaviour of these stars are similar.\\
In three stars of our N-rich sample we could estimate the ratio \Cdou/\Ctre~ (Fig.\,\ref{Fig:cdt}) and we found a ratio $\rm \leq 15$~ indicating a pollution by CNO processed material.

\subsection{Anti-correlation Na-O}
The main characteristic of the GC stars is the anti-correlation between the abundance of Na and the abundance of O \citep[see e.g.][]{CharbonnelCK14}.
In Fig.\,\ref{Fig:NaO} we present the relation [Na/Fe] vs. [O/Fe] for our sample of stars. We have over plotted the position of the giant stars of NGC6752 following \citet{BastianLardo18}, and of the dwarf stars following \citet{CarrettaGL05}.  The abundance of oxygen in dwarfs is estimated from the \Ou triplet, the lines are weak and a precise estimation requires a high S/N ratio. As a consequence the error in the determination of the oxygen abundance in the dwarfs of NGC6752 is large. We estimate that it is at least 0.2 dex and even 0.3 dex in particular for NGC6752-200613.\\ 
The general trend of our sample of stars agrees with the trend of the NGC\,6752 stars. 

This is the first time that the existence of a Na-O
anticorrelation has been demonstrated among field stars.
This has been possible because the N-rich stars of our sample
proved to be Na-rich and O-poor. This clearly indicates that
they are second generation GC stars. The corollary of this finding
is that first generation of GC stars have a chemical composition 
that is indistiguishable from that of field stars.

\subsection{Anti-correlation Li-Na}
A strong anti-correlation between Li and Na has been observed in NGC6752 \citep{PasquiniBM05}, but the slope of this anti-correlation is much weaker in M4 \citep{MonacoVB12,SpiteSG16} and in NGC\,6397 \citep{GonzalezHernandezBC09,LindPC09}. As a consequence, and since we never observe a very high excess of Na in our metal-poor N-rich stars we do not expect to observe a clear Li-Na anti-correlation, as it can be guessed from Fig.\,\ref{Fig:libe} where N-rich and normal reference stars seem to share the same Li abundance.

\subsection{Anti-correlation Al-Mg} 
In some GCs, an anti-correlation Al-Mg is observed \citep[][ and references therein]{MezarosMG20,BaezaFV22}. The Mg-Al cycle needs large temperatures (>70 million Kelvin) to operate but Al-rich stars, with [Al/Fe]>0.3, in GCs is sometimes observed in massive GCs.
G\,90-3 could be one of these stars.  Compared to the other stars, G\,90-3 is not Mg-poor, but since in the gas there is almost ten times more Mg than Al, the aluminium abundance is more sensitive to a pollution than Mg.

\subsection{The neutron-capture elements in N-rich stars and in the GCs. The [Eu/Ba] ratio.} \label{sec:heavyGC}
The heavy elements have been studied in detail in very few GCs and only in giant stars. However, since the abundance of these elements in the atmosphere, is not supposed to vary during the evolution of a star, at least during the main sequence and the red giant branch (RGB) phases, it is possible to compare the abundances of the heavy elements in dwarfs and in RGB stars.\\
 
In metal-poor GCs, like in very metal-poor stars), r-process production dominates, and the ratio [Eu/Ba] is generally close to 0.5\,dex as observed in the very metal-poor field stars, a value characteristic of the ``r'' process \citep[see e.g. ][]{FrancoisDH07,Spite-mf14}. 

 In a given cluster generally the abundance of the n-capture elements does not vary from star to star  \citep[see e.g.][]{Roederer11,KirbyDR20}. However a small number of GCs show variations, the best example of this phenomenon is M15 \citep{SobeckKS11,WorleyHS13}. \\
In this cluster, [Eu/Fe] and [Ba/Fe] varies and no correlation was established between [Na/Fe] and [Eu/Fe] or [Ba/Fe]. The large scatter of these abundances is attributed to an inhomogeneity of the original cloud. \\
In this cluster \citet{JohnsonBolte02} found that the ratio [Y/Zr] is lower than in the field stars. It seems also in Fig.\,\ref{abNHrich-YZrEuBa} (upper panel) that  [Y/Zr] is lower in the N-rich stars than in the normal stars. However the difference is not as large as the difference  (0.3\,dex) reported by \citet{JohnsonBolte02}.\\
Although [Eu/Fe] varies from star to star, in most of the M15 stars the ratio [Eu/Ba] is constant, close to the ``r'' process ratio (+0.5\,dex). However we remark that in this cluster, following the data of \citet{WorleyHS13}, there are two stars, in their ``S97'' sample, with a low ratio of [Eu/Ba] (high [Ba/Eu]), K969 and K479 \citep[Fig.\,13a in][]{WorleyHS13}. These stars  have also a rather high value of the Na abundance with $\rm[Na/Fe] \geq 0.5 dex$ \citep{SnedenKS97}.  Likewise, the star ID42262 in the ``W13'' sample with a very low [Eu/Ba] ratio \citep[Fig.\,13b in ][]{WorleyHS13} has the highest [Na/Fe] ratio of the sample ($\rm[Na/Fe]=+0.9 dex$). As a consequence, it seems that in GC stars, a high value of [Na/Fe] is sometimes (not always) associated with a low value of [Eu/Ba] indicating a ``s'' process contamination. 

One of our  N-rich stars, HD\,74000, with [Eu/Ba]=--0.05 at [Fe/H]=--2.0, seems to have a solar [Eu/Ba] ratio unexpected for its metallicity (see Fig. 10).  This star, like some Na-rich stars in M15, has been probably polluted by ``s'' process material ejected by a first generation AGB stars.

\section{Conclusion}  
We have studied a sample of eight dwarf stars suspected to be N-rich (and not enhanced in C), like the second generation stars of GCs.
The N abundance is deduced from the computation of the NH band in the near UV, based on the new parameters of \citet{FernandoBH18}.  
Two stars suspected to be N-rich by \citet{Laird85}, were found to have a normal N abundance. The six other stars, were found to have a ratio  [N/Fe]>1.0\,dex.

If the N-rich stars in the field, are second generation stars ejected from GCs, we do not expect the same homogeneity of their chemical composition as found in a given GC since these stars were formed in different clusters with different initial masses and thus different histories and metal-enrichment.
 
The abundances of C, N, O in our sample of N-rich stars present the same characteristics as those found in the the second generation stars of the GCs. C and O are slightly deficient but the scatter of [(C+N+O)/Fe] is very small. When [Fe/H] decreases from -0.5 to -2.0\,dex, [(C+N+O)/Fe] increases from about 0.2 to 0.6\,dex as observed in normal stars. This is a strong indication that the N enrichment is the result of a pollution by CNO processed material.

The N-rich stars of our sample, like the second generation stars in the GCs, show an excess of Na and sometimes of Al, as expected if these stars have been polluted by the ejecta of AGB stars. A Na-O anti-correlation is observed, similar to the relation observed in NGC6752.
 
The N-rich star HD\,74000 has a solar value of the ratio [Eu/Ba], this low value is unexpected for its metallicity ([Fe/H]=-2.0\,dex). At this metallicity indeed, the neutron-capture elements are generally only formed by the ``r'' process and [Eu/Ba]=+0.5\,dex.  We show that this peculiarity is also observed in at least three stars of M15 which all have an excess of the Na abundance, as expected for second generation stars. We suppose that in these stars a first generation of AGB stars has enriched the matter in ``s'' process material.

This analysis supports the hypothesis that the N-rich
stars, today observed in the field, were  formed as 
second generation stars in GCs and lost through the tidal
interaction between the GC and the Galaxy.

\begin {acknowledgements} 
This work uses results from the European Space Agency (ESA) space 
mission Gaia.  Gaia data are being processed by the Gaia Data
Processing and Analysis Consortium (DPAC).  Funding for the DPAC is
provided by national institutions, in particular the institutions
participating in the Gaia MultiLateral Agreement (MLA).  The Gaia
mission website is https://www.cosmos.esa.int/gaia.  The Gaia archive
website is https://archives.esac.esa.int/gaia.  
\end{acknowledgements}

\bibliographystyle{aa}

\begin{thebibliography}{}

\bibitem[Alvarez \& Plez(1998)]{AlvarezP98}
Alvarez R., Plez B., 1998, A\&A 330, 1109

\bibitem[Andrievsky et al.(2008)]{AndrievskySK08} 
Andrievsky, S.~M., Spite, M., Korotin, S.~A., et al.\ 2008, \aap, 481, 481

\bibitem[Arenou et al.(2018)]{ArenouLB18} 
Arenou, F., Luri, X., Babusiaux, C., et al.\ 2018, \aap, 616, A17

\bibitem[Baeza et al.(2022)]{BaezaFV22} 
Baeza, I., Fern{\'a}ndez-Trincado, J.~G., Villanova, S., et al.\ 2022, arXiv:2204.09703

\bibitem[Bastian \& Lardo(2018)]{BastianLardo18} 
Bastian, N. \& Lardo, C.\ 2018, \araa, 56, 83. doi:10.1146/annurev-astro-081817-051839

\bibitem[Bekki(2019)]{Bekki19} 
Bekki, K.\ 2019, \mnras, 490, 4007

\bibitem[Boesgaard(2007)]{Boesgaard07} 
Boesgaard, A.~M.\ 2007, \apj, 667, 1196. doi:10.1086/521075

\bibitem[Bonifacio et al.(2015)]{BonifacioCS15} 
Bonifacio, P., Caffau, E., Spite, M., et al.\ 2015, \aap, 579, A28. doi:10.1051/0004-6361/201425266

\bibitem[Bressan et al.(2012)]{BressanMG12} 
Bressan, A., Marigo, P., Girardi, L., et al.\ 2012, \mnras, 427, 127 

\bibitem[Carbon et al.(1986)]{CarbonKB86} 
Carbon, D.~F., Kraft, R.~P., Barbuy, B., et al.\ 1986, \rmxaa, 12, 173

\bibitem[Carretta et al.(2005)]{CarrettaGL05} 
Carretta, E., Gratton, R.~G., Lucatello, S., et al.\ 2005, \aap, 433, 597. doi:10.1051/0004-6361:20041892

\bibitem[Charbonnel et al.(2014)]{CharbonnelCK14} 
Charbonnel, C., Chantereau, W., Krause, M., et al.\ 2014, \aap, 569, L6

\bibitem[Choplin et al.(2017)]{ChoplinHM17} 
Choplin, A., Hirschi, R., Meynet, G., et al.\ 2017, \aap, 607, L3

\bibitem[Choplin et al.(2016)]{ChoplinMM16} 
Choplin, A., Maeder, A., Meynet, G., et al.\ 2016, \aap, 593, A36

\bibitem[Denissenkov \& Hartwick(2014)]{DenissenkovHartwick14} 
Denissenkov, P.~A. \& Hartwick, F.~D.~A.\ 2014, \mnras, 437, L21. doi:10.1093/mnrasl/slt133

\bibitem[Fernando et al.(2018)]{FernandoBH18} 
Fernando, A.~M., Bernath, P.~F., Hodges, J.~N., et al.\ 2018, \jqsrt, 217, 29. doi:10.1016/j.jqsrt.2018.05.021

\bibitem[Fran{\c{c}}ois et al.(2007)]{FrancoisDH07} 
Fran{\c{c}}ois, P., Depagne, E., Hill, V., et al.\ 2007, \aap, 476, 935

\bibitem[Gaia Collaboration et al.(2018)]{GaiaDR2-Brown18} 
Gaia Collaboration, Brown, A.~G.~A., Vallenari, A., et al.\ 2018, \aap, 616, A1 

\bibitem[Gieles \& Charbonnel(2020)]{GielesCharb19} 
Gieles, M. \& Charbonnel, C.\ 2020, Star Clusters: From the Milky Way to the Early Universe, 351, 297. doi:10.1017/S1743921319007658

\bibitem[Gonz{\'a}lez Hern{\'a}ndez et al.(2009)]{GonzalezHernandezBC09} 
Gonz{\'a}lez Hern{\'a}ndez, J.~I., Bonifacio, P., Caffau, E., et al.\ 2009, \aap, 505, L13. doi:10.1051/0004-6361/200912713

\bibitem[Gratton et al.(2000)]{GrattonSC00} 
Gratton, R.~G., Sneden, C., Carretta, E., et al.\ 2000, \aap, 354, 169

\bibitem[Gratton et al.(2019)]{GrattonBC19} 
Gratton, R., Bragaglia, A., Carretta, E., et al.\ 2019, \aapr, 27, 8. doi:10.1007/s00159-019-0119-3

\bibitem[Gruyters et al.(2014)]{GruytersNK14} 
Gruyters, P., Nordlander, T., \& Korn, A.~J.\ 2014, \aap, 567, A72. doi:10.1051/0004-6361/201423590

\bibitem[Gustafsson et al.(1975)]{GustafssonBE75}
Gustafsson B., Bell R. A., Eriksson K., Nordlund \AA., 1975, A\&A, 42, 407 

\bibitem[Gustafsson et al.(2003)]{GustafssonEE03}
Gustafsson B., Edvardsson B., Eriksson K., et al. 2003, in Stellar 
Atmosphere Modeling, ed. I. Hubeny, D. Mihalas, \& K. Werner, ASP Conf. Ser., 288, 
331 

\bibitem[Hansen et al.(2016b)]{HansenAN16b} 
Hansen, T.~T., Andersen, J., Nordstr{\"o}m, B., et al.\ 2016, \aap, 588, A3

\bibitem[Hansen et al.(2016a)]{HansenAN16a} 
Hansen, T.~T., Andersen, J., Nordstr{\"o}m, B., et al.\ 2016, \aap, 586, A160

\bibitem[Harmer \& Pagel(1973)]{HarmerPagel73} 
Harmer, D.~L. \& Pagel, B.~E.~J.\ 1973, \mnras, 165, 91. doi:10.1093/mnras/165.1.91

\bibitem[Ibata et al.(2021)]{IbataMM21} 
Ibata, R., Malhan, K., Martin, N., et al.\ 2021, \apj, 914, 123. doi:10.3847/1538-4357/abfcc2

\bibitem[Ishigaki et al.(2014)]{IshigakiTK14} 
Ishigaki, M.~N., Tominaga, N., Kobayashi, C., et al.\ 2014, \apjl, 792, L32. doi:10.1088/2041-8205/792/2/L32

\bibitem[Israelian et al.(2004)]{IsraelianER04} 
Israelian, G., Ecuvillon, A., Rebolo, R., et al.\ 2004, \aap, 421, 649. doi:10.1051/0004-6361:20047132

\bibitem[Johnson \& Bolte(2002)]{JohnsonBolte02} 
Johnson, J.~A. \& Bolte, M.\ 2002, \apj, 579, 616. doi:10.1086/342829

\bibitem[Kirby et al.(2020)]{KirbyDR20} 
Kirby, E.~N., Duggan, G., Ramirez-Ruiz, E., et al.\ 2020, \apjl, 891, L13. doi:10.3847/2041-8213/ab78a1

\bibitem[Laird(1985)]{Laird85} 
Laird, J.~B.\ 1985, \apj, 289, 556

\bibitem[Lallement et al.(2019)]{LallementBV18} 
Lallement, R., Babusiaux, C., Vergely, J.~L., et al.\ 2019, \aap, 625, A135

\bibitem[Lind et al.(2009)]{LindPC09} 
Lind, K., Primas, F., Charbonnel, C., et al.\ 2009, \aap, 503, 545

\bibitem[Lind et al.(2011)]{LindAB11} 
Lind, K., Asplund, M., Barklem, P.~S., et al.\ 2011, \aap, 528, A103. doi:10.1051/0004-6361/201016095

\bibitem[Lindegren et al.(2018)]{LindegrenHB18} 
Lindegren, L., Hern{\'a}ndez, J., Bombrun, A., et al.\ 2018, \aap, 616, A2

\bibitem[Lucatello et al.(2006)]{LucatelloBC06} 
Lucatello, S., Beers, T.~C., Christlieb, N., et al.\ 2006, \apjl, 652, L37

\bibitem[Maeder \& Meynet(2006)]{MaederMeynet06} 
Maeder, A. \& Meynet, G.\ 2006, \aap, 448, L37. doi:10.1051/0004-6361:200600012

\bibitem[Marigo et al.(2017)]{MarigoGB17} 
Marigo, P., Girardi, L., Bressan, A., et al.\ 2017, \apj, 835, 77

\bibitem[Martell et al.(2011)]{MartellSB11} 
Martell, S.~L., Smolinski, J.~P., Beers, T.~C., et al.\ 2011, \aap, 534, A136

\bibitem[Martin et al.(2022)]{MartinVA22} 
Martin, N.~F., Venn, K.~A., Aguado, D.~S., et al.\ 2022, \nat, 601, 45. doi:10.1038/s41586-021-04162-2

\bibitem[Mashonkina et al.(2010)]{MashonkinaCB10} 
Mashonkina, L., Christlieb, N., Barklem, P.~S., et al.\ 2010, \aap, 516, A46. doi:10.1051/0004-6361/200913825

\bibitem[Masseron et al.(2010)]{MasseronJP10} 
Masseron, T., Johnson, J.~A., Plez, B., et al.\ 2010, \aap, 509, A93

\bibitem[Masseron et al.(2014)]{MasseronPVE14} 
Masseron, T., Plez, B., Van Eck, S., et al.\ 2014, \aap, 571, A47

\bibitem[M{\'e}sz{\'a}ros et al.(2020)]{MezarosMG20} 
M{\'e}sz{\'a}ros, S., Masseron, T., Garc{\'\i}a-Hern{\'a}ndez, D.~A., et al.\ 2020, \mnras, 492, 1641. doi:10.1093/mnras/stz3496

\bibitem[Monaco et al.(2012)]{MonacoVB12} 
Monaco, L., Villanova, S., Bonifacio, P., et al.\ 2012, \aap, 539, A157. doi:10.1051/0004-6361/201117709

\bibitem[Mucciarelli et al.(2019)]{MucciarelliLL19} 
Mucciarelli, A., Lapenna, E., Lardo, C., et al.\ 2019, \apj, 870, 124. doi:10.3847/1538-4357/aaf3a4

\bibitem[Nordlander \& Lind(2017)]{NordlanderLind17} 
Nordlander, T. \& Lind, K.\ 2017, \aap, 607, A75. doi:10.1051/0004-6361/201730427

\bibitem[Pace(2013)]{Pace13} 
Pace, G.\ 2013, \aap, 551, L8. doi:10.1051/0004-6361/201220364

\bibitem[Peterson et al.(2020)]{PetersonBS20} 
Peterson, R.~C., Barbuy, B., \& Spite, M.\ 2020, \aap, 638, A64

\bibitem[Pasquini et al.(2004)]{PasquiniBR04} 
Pasquini, L., Bonifacio, P., Randich, S., et al.\ 2004, \aap, 426, 651. doi:10.1051/0004-6361:20041254

\bibitem[Pasquini et al.(2005)]{PasquiniBM05} 
Pasquini, L., Bonifacio, P., Molaro, P., et al.\ 2005, \aap, 441, 549

\bibitem[Pasquini et al.(2007)]{PasquiniBR07} 
Pasquini, L., Bonifacio, P., Randich, S., et al.\ 2007, \aap, 464, 601

\bibitem[Pasquini et al.(2008)]{PasquiniEB08} 
Pasquini, L., Ecuvillon, A., Bonifacio, P., et al.\ 2008, \aap, 489, 315

\bibitem[Peterson et al.(2020)]{PetersonBS20} 
Peterson, R.~C., Barbuy, B., \& Spite, M.\ 2020, \aap, 638, A64. doi:10.1051/0004-6361/202037689

\bibitem[Plez(2008)]{Plez08} 
Plez, B.\ 2008, Physica Scripta Volume T, 133, 014003. doi:10.1088/0031-8949/2008/T133/014003

\bibitem[Plez(2012)]{Plez-code12} 
Plez, B.\ 2012, Turbospectrum: Code for spectral synthesis, ascl:1205.004

\bibitem[Prantzos et al.(2007)]{PrantzosCI07} 
Prantzos, N., Charbonnel, C., \& Iliadis, C.\ 2007, \aap, 470, 179

\bibitem[Ram{\'\i}rez et al.(2012)]{RamirezMC12} 
Ram{\'\i}rez, I., Mel{\'e}ndez, J., \& Chanam{\'e}, J.\ 2012, \apj, 757, 164

\bibitem[Rich \& Boesgaard(2009)]{RichBoesgaard09} 
Rich, J.~A. \& Boesgaard, A.~M.\ 2009, \apj, 701, 1519

\bibitem[Roederer et al.(2010)]{RoedererCK10} 
Roederer, I.~U., Cowan, J.~J., Karakas, A.~I., et al.\ 2010, \apj, 724, 975. doi:10.1088/0004-637X/724/2/975

\bibitem[Roederer(2011)]{Roederer11} 
Roederer, I.~U.\ 2011, \apjl, 732, L17. doi:10.1088/2041-8205/732/1/L17

\bibitem[Roederer \& Lawler(2012)]{RoedererLawler12} 
Roederer, I.~U. \& Lawler, J.~E.\ 2012, \apj, 750, 76

\bibitem[Schiappacasse-Ulloa et al.(2022)]{SchiappacasseLR22} 
Schiappacasse-Ulloa, J., Lucatello, S., Rain, M.~J., et al.\ 2022, \mnras, 511, 231

\bibitem[Schiavon et al.(2017)]{SchiavonJF17} 
Schiavon, R.~P., Johnson, J.~A., Frinchaboy, P.~M., et al.\ 2017, \mnras, 466, 1010

\bibitem[Schlegel et al.(1998)]{SchlegelFD98} 
Schlegel, D.~J., Finkbeiner, D.~P., \& Davis, M.\ 1998, \apj, 500, 525

\bibitem[Schuster(1981)]{Schuster81} 
Schuster, W.~J.\ 1981, \rmxaa, 5, 69

\bibitem[Siqueira-Mello et al.(2015)]{SiqueiraAB15} 
Siqueira-Mello, C., Andrievsky, S.~M., Barbuy, B., et al.\ 2015, \aap, 584, A86

\bibitem[Sneden et al.(1997)]{SnedenKS97} 
Sneden, C., Kraft, R.~P., Shetrone, M.~D., et al.\ 1997, \aj, 114, 1964. doi:10.1086/118618

\bibitem[Sobeck et al.(2011)]{SobeckKS11} 
Sobeck, J.~S., Kraft, R.~P., Sneden, C., et al.\ 2011, \aj, 141, 175. doi:10.1088/0004-6256/141/6/175

\bibitem[Spite \& Spite(1986)]{SpiteSpite86} 
Spite, F., \& Spite, M.\ 1986, \aap, 163, 140

\bibitem[Spite et al.(2006)]{SpiteCH06}
Spite, M., Cayrel, R., Hill, V., et al.\ 2006, \aap, 455, 291

\bibitem[Spite et al.(2005)]{SpiteCP05} 
Spite, M., Cayrel, R., Plez, B., et al.\ 2005, \aap, 430, 655

\bibitem[Spite \& Spite(2014)]{Spite-mf14} 
Spite, M. \& Spite, F.\ 2014, Astronomische Nachrichten, 335, 65. doi:10.1002/asna.201311998

\bibitem[Spite et al.(2016)]{SpiteSG16} 
Spite, M., Spite, F., Gallagher, A.~J., et al.\ 2016, \aap, 594, A79. doi:10.1051/0004-6361/201628759

\bibitem[Tang et al.(2020)]{TangFTL20} 
Tang, B., Fern{\'a}ndez-Trincado, J.~G., Liu, C., et al.\ 2020, \apj, 891, 28

\bibitem[Worley et al.(2013)]{WorleyHS13} 
Worley, C.~C., Hill, V., Sobeck, J., et al.\ 2013, \aap, 553, A47. doi:10.1051/0004-6361/201321097

\bibitem[Yuan et al.(2022)]{YuanMI22} 
Yuan, Z., Martin, N.~F., Ibata, R.~A., et al.\ 2022, MNRAS, submitted, arXiv:2203.02512

\bibitem[Zhao et al.(2016)]{ZhaoMY16} 
Zhao, G., Mashonkina, L., Yan, H.~L., et al.\ 2016, \apj, 833, 225

\end{thebibliography}
{}

\end{document}